\documentclass[aps,prd,preprint,groupedaddress,nofootinbib,superscriptaddress,floatfix,11pt]{revtex4-1}

\usepackage{amsfonts}
\usepackage{amsmath}
\usepackage{bm,graphicx}
\usepackage{hyperref}
\usepackage{epsfig}
\usepackage{verbatim}   
\usepackage{epstopdf}
\usepackage{dcolumn}
\usepackage{color}
\usepackage{tabularx}
\usepackage{braket}
\usepackage{tikz}
\usepackage{amssymb}
\usepackage[percent]{overpic}
\graphicspath{{./figures/}}
\usepackage{subfigure}
\usepackage{mwe}
\usepackage{soul}
\usepackage{ulem}
\usepackage{comment}
\usepackage[section]{placeins}

\hypersetup{colorlinks, linkcolor = [rgb]{0,0.0,0.5}, citecolor = [rgb]{0,0.0,0.5}, urlcolor = [rgb]{0,0.0,0.5}}

\usepackage[utf8]{inputenc}
\usepackage{lineno}
\def\be{\begin{equation}}
\def\ee{\end{equation}}
\def\bea{\begin{eqnarray}}
\def\eea{\end{eqnarray}}

\newcommand{\nn}{\nonumber}
\newcommand{\la}{\label}

\newcommand{\half}{\textstyle \frac{1}{2}}

\definecolor{green}{rgb}{0,.5,0}

\begin{document}


\preprint{JLAB-THY-18-2803}
\preprint{SLAC-PUB-17327}

\title{Nonperturbative strange-quark sea from lattice QCD, 
light-front holography, and meson-baryon fluctuation models}

\newcommand*{\JLAB}{Thomas Jefferson National Accelerator Facility, Newport News, VA 23606, USA}\affiliation{\JLAB}
\newcommand*{\DUKE}{Department of Physics, Duke University, Durham, North Carolina 27708, USA}\affiliation{\DUKE}
\newcommand*{\COSTARICA}{Laboratorio de F\'isica Te\'orica y Computacional, Universidad de Costa Rica, 11501 San Jos\'e, Costa Rica}\affiliation{\COSTARICA}
\newcommand*{\HEIDELBERG}{Institut f\"ur Theoretische Physik der Universit\"at, D-69120 Heidelberg, Germany}\affiliation{\HEIDELBERG}
\newcommand*{\SLAC}{SLAC National Accelerator Laboratory, Stanford University, Stanford, CA 94309, USA}\affiliation{\SLAC}
\newcommand*{\SCSU}{Computer Science Department, Southern Connecticut State University, New Haven, CT 06515, USA}\affiliation{\SCSU}
\newcommand*{\PKU}{School of Physics and State Key Laboratory of Nuclear Physics and Technology, Peking University, Beijing 100871, China}\affiliation{\PKU}
\newcommand*{\CIC}{Collaborative Innovation Center of Quantum Matter, Beijing, China}\affiliation{\CIC}
\newcommand*{\CHEP}{Center for High Energy Physics, Peking University, Beijing 100871, China}\affiliation{\CHEP}

\author{Raza~Sabbir~Sufian}\affiliation{\JLAB}
\author{Tianbo~Liu}\email{liutb@jlab.org}\affiliation{\JLAB}\affiliation{\DUKE}
\author{Guy~F.~de~T\'eramond}\affiliation{\COSTARICA}
\author{Hans G\"unter Dosch}\affiliation{\HEIDELBERG}
\author{Stanley~J.~Brodsky}\affiliation{\SLAC}
\author{Alexandre~Deur}\affiliation{\JLAB}
\author{Mohammad~T.~Islam}\affiliation{\SCSU}
\author{Bo-Qiang~Ma}\affiliation{\PKU}\affiliation{\CIC}\affiliation{\CHEP}

\collaboration{HLFHS Collaboration}\noaffiliation

\begin{abstract}

We demonstrate that a nonzero strangeness contribution to the spacelike electromagnetic form factor of the nucleon  is evidence for a strange-antistrange asymmetry in the nucleon's light-front wave function,  thus implying different nonperturbative contributions to the strange and antistrange quark distribution functions. A recent lattice QCD calculation of the nucleon strange quark form factor predicts that the strange quark distribution is more centralized in coordinate space than the antistrange quark distribution, and thus the strange quark distribution is more spread out in light-front momentum space.  We show that the lattice prediction implies that the difference between the strange and antistrange parton distribution functions, $s(x)-\bar{s}(x)$, is negative at small-$x$ and positive at large-$x$.  We also evaluate the strange quark form factor and $s(x)-\bar{s}(x)$  using a baryon-meson fluctuation model and a novel nonperturbative model based on light-front holographic QCD. This procedure leads to a Veneziano-like expression of the form factor, which depends exclusively on the twist of the hadron and the properties of the Regge trajectory of the vector meson which couples to the quark current in the hadron. The holographic structure of the model allows us to introduce unambiguously quark masses in the form factors and quark distributions preserving the hard scattering counting rule at large-$Q^2$ and the inclusive counting rule at large-$x$. Quark masses modify the Regge intercept which governs the small-$x$ behavior of quark distributions, therefore modifying their small-$x$ singular behavior. Both nonperturbative approaches provide descriptions of the strange-antistrange asymmetry and intrinsic strangeness in the nucleon consistent with the lattice QCD result. 

\end{abstract}

\maketitle

\section{\la{intro} Introduction}

The unveiling of nucleon structure in terms of fundamental quark and gluonic degrees of freedom is a main goal of nuclear and particle physics. The strangeness distribution of the nucleon is of particular interest  since it is a purely sea-quark distribution.  The nonperturbative dynamics of the strange-antistrange quark asymmetry $s(x)-\bar{s}(x)$ poses a challenging theoretical problem. It  has become of major interest in both experimental and phenomenological studies, not only because of its important role in understanding strong-interaction dynamics but also because  the $s(x)-\bar{s}(x)$ asymmetry is an important  input for testing  electroweak theory and new physics models.  For example, a precise test of electroweak physics in neutrino and antineutrino-induced dimuon production depends in detail on the intrinsic strange and antistrange distributions in the nucleon~\cite{Musolf:1993tb}.  The intrinsic nonperturbative strangeness distributions and asymmetry also give insight, via the operator product expansion, into the nonperturbative physics of the intrinsic charm and bottom contributions to the nucleon structure functions~\cite{Brodsky:1984nx,Brodsky:1980pb,Franz:2000ee}.

Lattice QCD calculations~\cite{Sufian:2016pex,Sufian:2016vso,Sufian:2017osl}, at the physical pion mass and extrapolated to the continuum and infinite volume limits,  have provided  estimates  of the strangeness  contribution to the electromagnetic (EM) form factors of the nucleon with better accuracy than that available from the global analyses~\cite{Liu:2007yi,Gonzalez-Jimenez:2014bia,Young:2006jc} of the experimental data.  A direct lattice calculation of $s(x)-\bar{s}(x)$ has not as yet been achieved~\cite{Lin:2017snn}. However, we shall show that  one can constrain the $s(x)-\bar{s}(x)$ asymmetry by comparing the lattice QCD results for the strange form factor with predictions based on a baryon-meson fluctuation model~\cite{Brodsky:1996hc}.  We will also introduce a new model based on the structural behavior of the light-front holographic approach to hadron structure~\cite{Brodsky:2014yha}, form factors and parton distribution functions~\cite{deTeramond:2018ecg}. We shall show that the $s(x)-\bar{s}(x)$  asymmetry in the nucleon can be predicted up to a normalization factor constrained by lattice results.

Parton distribution functions (PDFs) are interpreted, at leading twist, as  distributions of quarks and gluons carrying the light-front momentum fraction $x$ of the nucleon's momentum at fixed light-front time $\tau=t+z/c$.  The global QCD analysis of PDFs is based on factorization theorems of physical observables, such as the cross section of deep inelastic lepton-nucleon scattering~\cite{Collins:2011zzd}.  Although equal numbers of $s$ and $\bar{s}$ are required by their nonvalence nature in the nucleon, 
\begin{align}
\langle s-\bar{s}\rangle = \int_0^1 dx\, \big[s(x) - \bar{s}(x)\big] = 0,\label{firstmoment}
\end{align}
no fundamental principles prohibit different $s(x)$ and $\bar{s}(x)$ distributions. A nonzero $s(x)-\bar{s}(x)$  has also been allowed for in  global analyses of PDFs~\cite{Ball:2014uwa,Harland-Lang:2014zoa,Jimenez-Delgado:2014twa}. Furthermore, the first moment of the difference of PDFs,
\begin{align} \label{Sm}
\langle S_- \rangle \equiv \big\langle x \big(s - \bar{s}\big) \big\rangle= \int_0^1 dx \, x  \big[s(x) - \bar{s}(x)\big],
\end{align}
can also be used to quantify the  $s(x)-\bar{s}(x)$ asymmetry.

The strange-quark sea in the nucleon has both ``extrinsic" and ``intrinsic" components~\cite{Brodsky:1984nx,Brodsky:1980pb,Franz:2000ee}. The extrinsic one is produced by gluon splitting $g\to s\bar{s}$ triggered by a hard probe, {\it e.g.,} the virtual photon exchanged between the lepton and the nucleon in a  deep inelastic scattering process. Since the QCD coupling $\alpha_s$ is small at high momentum scale, the extrinsic strange-sea derived from the splitting function can be calculated perturbatively. The nonperturbative intrinsic strange-sea  encoded in the nucleon's nonvalence  light-front (LF) Fock state wave function can in principle be obtained by solving the LF Hamiltonian eigenvalue problem~\cite{Pauli:1985ps}; {\it e.g.}, by matrix diagonalization. However,  to capture the nonperturbative dynamics in the bound state equations, one should integrate out all higher Fock states, corresponding to an infinite number of degrees of freedom, a formidable problem.

The strange-antistrange asymmetry in the nucleon originates in QCD from the difference between quark-quark versus quark-antiquark interactions. Since the nucleon carries nonzero quark number---the number of quarks minus the number of antiquarks---the interaction of the strange quark with the spectators of the nonvalence Fock states is different from that of the antistrange quark with the remaining quarks, thus leading to different $s$ and $\bar{s}$ distributions. The extrinsic strange-antistrange asymmetry in the nucleon PDF  arises from perturbative QCD evolution at high orders due to the difference between quark-to-strange quark splitting function $P_{qs}$ and quark-to-antistrange quark splitting function $P_{q\bar{s}}$.  Since the strange-antistrange pair is generated from a nonstrange quark at next-to-leading order, and the interaction between the strange/antistrange quark and the nonstrange quark is mediated by additional gluon exchange, this pQCD effect arises at the three-loop level.  An explicit calculation has been performed in~\cite{Catani:2004nc}.

In addition to PDFs, one can also obtain information on nucleon structure from elastic form factors, which relate to the transverse coordinate space distributions at fixed LF time via a Fourier transform~\cite{Miller:2010nz}.  The nucleon spin-preserving amplitude is described by the Dirac form factor, which can be expressed as:
\begin{align}\label{flavordecom}
F_1(Q^2)=\sum_q e_q F_1^q(Q^2),
\end{align}
where $Q^2$ is the momentum transfer squared, and the flavor form factor $F_1^q(Q^2)$, with $q=u,d,s,\cdots,$ measures the $q$-flavor quark contribution minus the $\bar{q}$-flavor antiquark contribution due to the opposite charge of the quark and antiquark. Therefore a nonvanishing $F_1^s(Q^2)$ at $Q^2 \ne 0$ indicates a strange-antistrange asymmetry in LF coordinate space. The constraint $F_1^s(0)=0$ is fixed by the sum rule~\eqref{firstmoment}.

Lattice QCD results for $F_1^s(Q^2)$, obtained in the continuum limit~\cite{Sufian:2016pex,Sufian:2016vso,Sufian:2017osl}, are shown in Fig.~\ref{f1sWF} with systematic and statistical uncertainties added in quadrature. The lattice QCD analyses are described in the Appendix~\ref{LQCD}.

There have been a number of phenomenological studies~\cite{Signal:1987gz,Cao:1999da,Brodsky:1996hc,Cao:1999fs,Melnitchouk:1999mv,Cao:2003ny,Feng:2012gu} of the $s(x)-\bar{s}(x)$ distribution. In the baryon-meson fluctuation model~\cite{Brodsky:1996hc}, the nonperturbative strange sea is generated from the fluctuation of the nucleon valence state to the lightest mass hadronic state with strangeness; {\it i.e.}, a kaon and a hyperon $(\Lambda\text{ or }\Sigma)$. The different distributions of the strange quark in the hyperon and the antistrange quark in the kaon yield a nonvanishing $s(x)-\bar{s}(x)$ distribution.  
In this model, a meson-baryon configuration, {\it e.g.}, the $K^+\Lambda^0$ state, creates different radially separated distributions of the $s$ and the $\bar{s}$ quarks from the center of mass.  Since the kaon is lighter than the hyperon, one expects that the kaon---and thus the $\bar{s}$ quark---to be at a larger radial distance from the center of mass than the hyperon and its $s$ quark.  This picture leads to $F_1^s(Q^2)>0$ at $Q^2>0$, consistent with the lattice QCD results~\cite{Sufian:2016pex,Sufian:2016vso,Sufian:2017osl}.

As we will discuss below,  a positive value of  $F_1^s(Q^2)$ at $Q^2>0$ indicates that the strange quark distribution is more centralized in coordinate space than the antistrange quark distribution, and results in an $s(x)-\bar{s}(x)$ asymmetry in momentum space.  A narrower distribution in coordinate space corresponds to a wider one in momentum space, and therefore the lattice QCD result $F_1^s(Q^2)>0$ implies a negative $s(x)-\bar{s}(x)$ distribution at small-$x$ and a positive distribution at large-$x$.

We will also examine in this article the behavior of $F_1^s(Q^2)$ and $s(x)-\bar{s}(x)$ using the nonperturbative structure of light-front holographic QCD (LFHQCD), a semiclassical  approach to relativistic bound state equations which follows from the holographic embedding of light-front dynamics in a higher dimensional gravity theory, with the constraints imposed by the underlying superconformal algebraic structure~\cite{Brodsky:2006uqa, deTeramond:2008ht, deTeramond:2013it, deTeramond:2014asa, Dosch:2015nwa, Brodsky:2016yod}. This approach incorporates a nontrivial connection to the hadron spectrum and therefore to the Regge trajectories predicted by the model.

In Sec.~\ref{General}, we will describe  the strange-antistrange asymmetries in coordinate and momentum spaces in the boost invariant light-front formalism, together with qualitative discussions.  We will perform quantitative calculations of  $s(x)-\bar{s}(x)$ and $F_1^s(Q^2)$   in Sec.~\ref{FluctModel} using the baryon-meson fluctuation model, and in Sec.~\ref{HologModel} using the structural framework of LFHQCD.   We will analyze the constraints imposed from lattice QCD for these two nonperturbative models. We will also use the lattice QCD data to quantitatively constrain each model in order to obtain more precise predictions.  The procedures discussed here can be applied to other approaches, {\it e.g.}, by deriving constraints on the wave functions predicted by meson cloud and chiral quark models.  Final discussions and conclusions are presented in Sec.~\ref{Concl}.


\section{\la{General} Strange-antistrange asymmetry in the nucleon}

Hadrons are eigenstates of the QCD LF Hamiltonian $H^{\rm QCD}_{\rm LF} |\Psi \rangle = M^2 |\Psi \rangle$~\cite{Brodsky:1997de}.  The hadronic light-front wave functions are the projection of  the eigenstate on the basis of free Fock states.
Taking a complete basis of LF Fock states with quarks and gluons as the degrees of freedom, a nucleon state with four-momentum 
$P^\mu = \left(P^+, P^-, {\bf P}_\perp \right)$ and total spin $S^z$ can be expanded as
\begin{align}\label{fockexp}
|N;P^+, {\bf P}_\perp, S^z\rangle &=\sum_{n,\{\lambda_i\}}\int[dx][d^2{\bf k}_{\perp}]\psi_{n/N}(x_i,{\bf k}_{i\perp},\lambda_i)|n;  x_i P^+,  x_i  {\bf P}_\perp + {\bf k}_{i\perp},\lambda_i\rangle,
\end{align}
where
\begin{align}
[dx][d^2{\bf k}_{\perp}]=16\pi^3\delta\Big(1-\sum_j x_j\Big)\delta^{(2)}\Big(\sum_j {\bf k}_{j\perp}\Big)\prod_{i}\frac{dx_id^2{\bf k}_{i\perp}}{2\sqrt{x_i}(2\pi)^3}.
\end{align}
The index $n=qqq, qqqg, qqqq\bar{q},\cdots,$ represents the constituents of the Fock state, the internal LF variables $x_i$, ${\bf k}_{i\perp}$, and $\lambda_i$ are the longitudinal momentum fraction, the intrinsic transverse momentum, and the spin carried by the  $i$th constituent, respectively, and $\psi_{n/N}$ is the light-front wave function (LFWF). It gives the probability  of the $n$-particle LF Fock state and represents the transition amplitude of the on-shell nucleon eigenstate to the quark and gluon Fock states of the free LF Hamiltonian which are off-shell in invariant mass. All nucleon properties are encoded in the LFWFs, which in principle could be obtained by solving the LF Hamiltonian eigenvalue problem. Aiming at a first-principle calculation of the LFWFs, calculational methods based on matrix diagonalization, such as discretized LF quantization~\cite{Pauli:1985pv},  the transverse lattice method~\cite{Bardeen:1976tm}  and the basis LF quantization~\cite{Vary:2009gt}, have been proposed.

In this paper, we will focus on the $s$ and $\bar{s}$ quark contributions to the nucleon nonvalence LF Fock state wave functions, $\psi_{s/N}(x_s,{\bf k}_{s\perp},\lambda_s)$ and $\psi_{\bar{s}/N}(x_{\bar{s}},{\bf k}_{\bar{s}\perp},\lambda_{\bar{s}})$, where the sum over other degrees of freedom is implied. The $s$ and $\bar{s}$ quark PDFs  expressed in terms of the LFWFs  are
\begin{align}
s(x)&=\sum_{\lambda_s}\int\frac{d^2{\bf k}_{s\perp}}{16\pi^3}|\psi_{s/N}(x_s,{\bf k}_{s\perp},\lambda_s)|^2,\\
\bar{s}(x)&=\sum_{\lambda_{\bar{s}}}\int\frac{d^2{\bf k}_{\bar{s}\perp}}{16\pi^3}|\psi_{\bar{s}/N}(x_{\bar{s}},{\bf k}_{\bar{s}\perp},\lambda_{\bar{s}})|^2.
\end{align}
The sum rule~\eqref{firstmoment} requires the normalization
\begin{align}\label{snorm}
\sum_{\lambda_s}\int\frac{dx_s d^2{\bf k}_{s\perp}}{16\pi^3}|\psi_{s/N}(x_s,{\bf k}_{s\perp},\lambda_s)|^2=\sum_{\lambda_{\bar{s}}}\int\frac{dx_{\bar{s}}d^2{\bf k}_{\bar{s}\perp}}{16\pi^3}|\psi_{\bar{s}/N}(x_{\bar{s}},{\bf k}_{\bar{s}\perp},\lambda_{\bar{s}})|^2= I_s,
\end{align}
where $I_s$ gives the number of intrinsic strange/antistrange quarks in the nucleon. Perturbative QCD evolution needs to be performed to include contributions from the extrinsic sea and to compare with the PDFs extracted from high energy scattering experiments.

The EM form factors of the nucleon are defined as~\cite{Foldy:1952}
\begin{align}
\langle P',S'|J^\mu(0)|P,S\rangle&=\bar{u}(P',S')\big[\gamma^\mu F_1(Q^2)
+\frac{i\sigma^{\mu\nu}q_\nu}{2M}F_2(Q^2)\big]u(P,S),
\end{align}
where $J^\mu=\sum_q e_q\bar{\psi}_q\gamma^\mu\psi_q$ is the current operator, $M$ is the nucleon mass, $F_1(Q^2)$ and $F_2(Q^2)$ are the Dirac and Pauli form factors, respectively.
Comparing with the decomposition \eqref{flavordecom}, one observes that $F_1^s(Q^2)$ and $F_2^s(Q^2)$ are given by the matrix elements of the current operator $J_s^\mu = \bar{\psi}_s\gamma^\mu\psi_s$.   In the LF formalism, $F_1(Q^2)$ and $F_2(Q^2)$ can be calculated from the overlap of spin-conserving and spin-flip matrix elements of the $+$ component of the current, respectively, ~\cite{Brodsky:1980zm}:
\begin{align}
\Big\langle P',\uparrow\Big| \frac{J^+(0)}{2P^+}\Big|P,\uparrow\Big\rangle&=F_1(q^2),\label{f1calc}\\
\Big\langle P',\uparrow\Big| \frac{J^+(0)}{2P^+}\Big|P,\downarrow\Big\rangle&=-\frac{q_1-iq_2}{2M}F_2(q^2),\label{f2calc}
\end{align}
with $q^\mu = \left(q^+,q^-, \mathbf{q_\perp} \right)$ and transferred momentum squared $q^2=t=(P'-P)^2 =-Q^2$.

The Drell-Yan-West (DYW) frame~\cite{Drell:1969km,West:1970av}
\begin{align}
q &=\Big(0,\frac{q^2}{2P^+},\mathbf{q}_\perp\Big),\\
P&=\Big(P^+, \frac{M^2}{2P^+}, {\bf 0}_\perp\Big),
\end{align}
with $q^2 = - \mathbf{q}_\perp^2$, can be used to avoid off-diagonal contributions  $n \to n'= n \pm 2 $ from Fock states with different constituents. Here ${\bf q}_\perp$ is the Fourier conjugate of the transverse LF coordinate ${\bf a}_\perp$.  From \eqref{f1calc} and the Fock state expansion \eqref{fockexp}, the Dirac form factor{,} in terms of the LFWFs{, is given by} the DYW expression~\cite{Drell:1969km,West:1970av}
\begin{align} \label{F1psi}
F_1^s(Q^2={\bf q}_{\perp}^2)&=\sum_{\lambda_s}\int\frac{dx_s d^2{\bf k}_{s\perp}}{16\pi^3}\psi_{s/N}^*(x_s,{\bf k}_{s\perp}+(1-x_s){\bf q}_{\perp},\lambda_s)\psi_{s/N}(x_s,{\bf k}_{s\perp},\lambda_s)\nn\\
&\quad-\sum_{\lambda_{\bar{s}}}\int\frac{dx_{\bar{s}}d^2{\bf k}_{\bar{s}\perp}}{16\pi^3}\psi_{\bar{s}/N}^*(x_{\bar{s}},{\bf k}_{\bar{s}\perp}+(1-x_{\bar{s}}){\bf q}_\perp,\lambda_{\bar{s}})\psi_{\bar{s}/N}(x_{\bar{s}},{\bf k}_{\bar{s}\perp},\lambda_{\bar{s}})\\
&=\rho_s({\bf q}_\perp)-\rho_{\bar{s}}({\bf q}_\perp), \nn
\end{align}
where $\rho_{s/\bar s}({\bf q}_\perp)$ represents the effective strange/antistrange density.  The relative minus sign  in (\ref{F1psi}) arises from the opposite strange and antistrange charges.

The density $\rho_{s/\bar s}({\bf q}_\perp)$  is the inverse Fourier transform of the distribution $\tilde{\rho}({\bf a}_\perp)$,
\begin{align}
\rho_{s/\bar{s}}({\bf q}_\perp)&=\int\frac{d^2{\bf a}_\perp}{(2\pi)^2}e^{i{\bf q}_\perp\cdot {\bf a}_\perp}\tilde{\rho}_{s/\bar{s}}({\bf a}_\perp).
\end{align}
Following the normalization~\eqref{snorm} or the sum rule~\eqref{firstmoment}, we require 
\begin{align}
\int d^2{\bf a}_\perp\tilde{\rho}_s({\bf a}_\perp)=\int d^2{\bf a}_\perp\tilde{\rho}_{\bar{s}}({\bf a}_\perp)= I_s,
\end{align}
and thus
$F_1^s(0)=0$.

\begin{figure}[htp]
\centering
\includegraphics[width=0.48\textwidth]{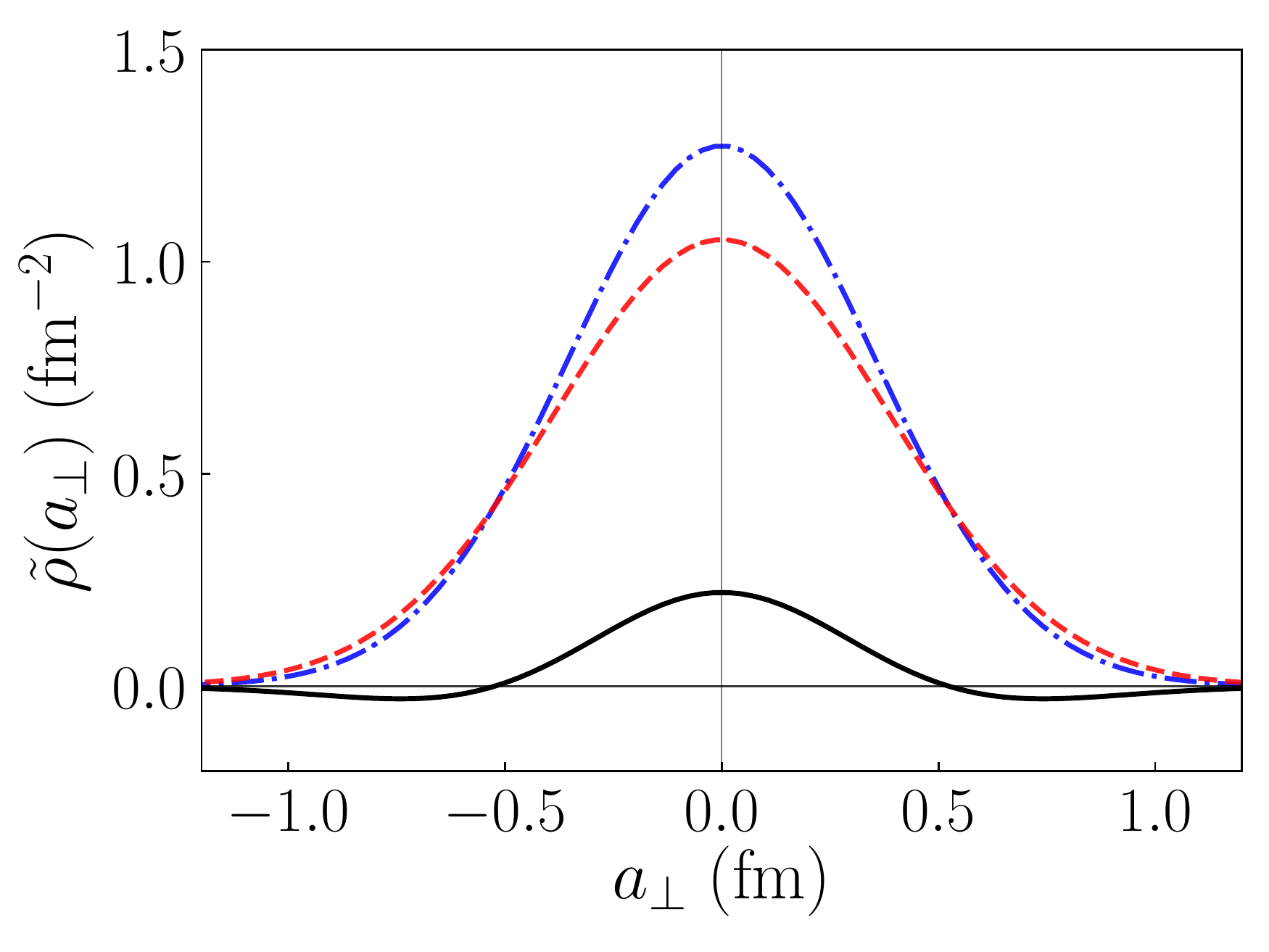}
\includegraphics[width=0.48\textwidth]{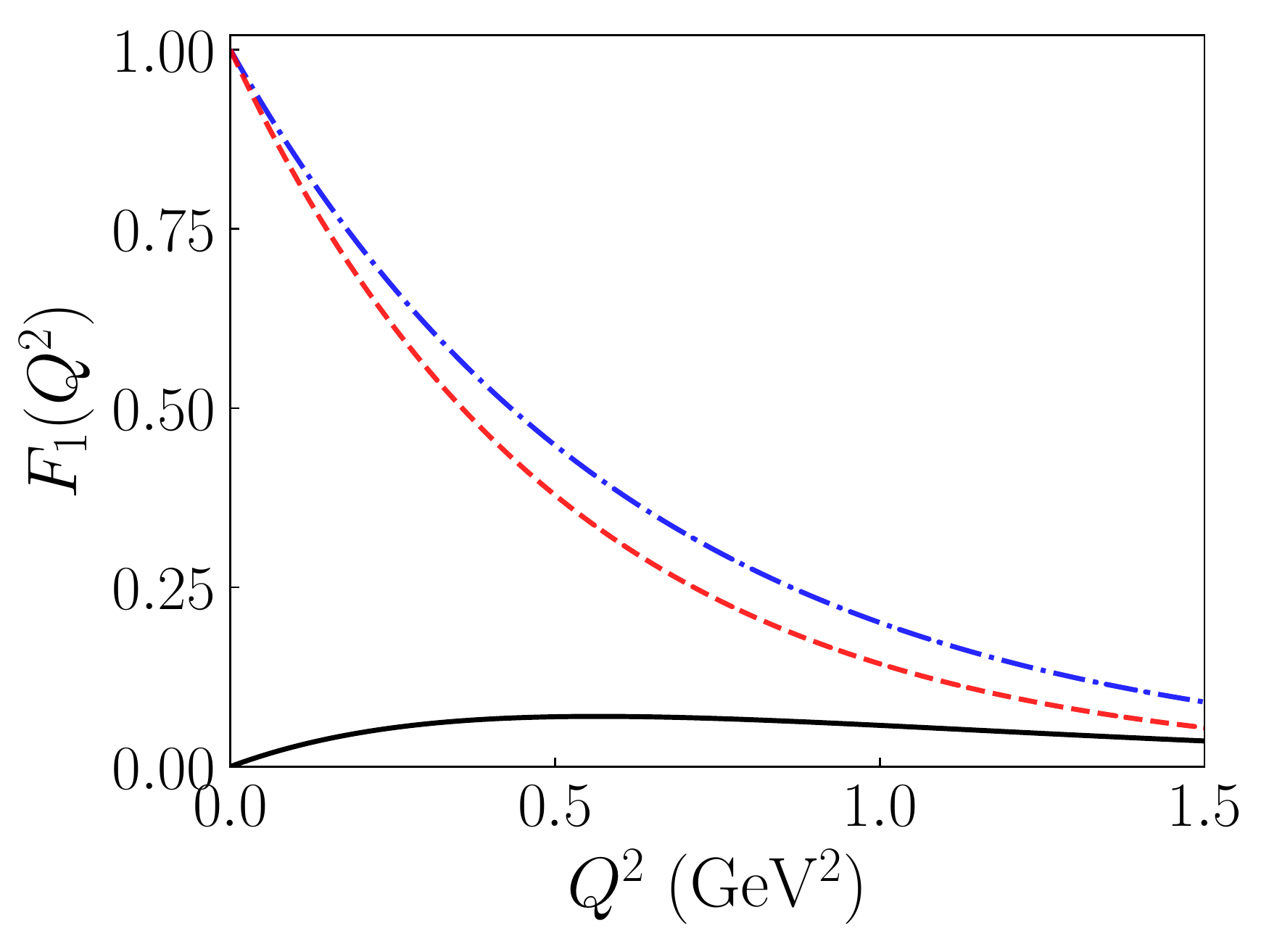}
\caption{Nonzero form factor $F_1(Q^2)$ (right panel) from asymmetric sea quark and antiquark distributions in transverse LF coordinate space (left panel). The dashed-dotted curves (blue) represent the quark, the dashed curves (red) represent the antiquark, and the continuous curves (black) represent $q-\bar{q}$. The quark/antiquark number is normalized to $1$ in this figure. \label{f1inter} }
\end{figure}

 A nonzero $F_1^s(Q^2)$  is equivalent to an asymmetric distribution $\tilde{\rho}_s({\bf a}_\perp)\neq \tilde{\rho}_s({\bf a}_\perp)$ based on the uniqueness of the Fourier transform.  As illustrated in Fig.~\ref{f1inter} for a simple Gaussian distribution, if the $s$ (or $\bar{s}$) quarks are more concentrated at small transverse separation than the  $\bar{s}$ (or $s$) quarks, one obtains a positive (or negative) form factor $F_1^s(Q^2)$ at $Q^2>0$.  A similar concept based on the locality defined in the instant form was presented in~\cite{Ji:1995rd}.

The strange-antistrange asymmetries in LF coordinate space and LF momentum space are correlated.  To show this, we express $\rho_{s/\bar{s}}({\bf q}_\perp)$ in terms of the transverse impact variable $\bf{b}_\perp$ using the Fourier transform of the ${\bf k}_\perp$-space LFWFs following Ref.~\cite{Soper:1976jc},
\begin{align}
\rho_{s/\bar{s}}({\bf q}_\perp)&=\sum_{\lambda_{s/\bar{s}}}\int dx_{s/\bar{s}} \, d^2{\bf b}_\perp \exp\Big(i(1-x_{s/\bar{s}}){\bf b}_\perp\cdot {\bf q}_\perp\Big) \left\vert \tilde{\psi}_{s/\bar{s}}(x_{s/\bar{s}},{\bf b}_\perp, \lambda_{s/\bar{s}})\right\vert^2.
\end{align}
The coordinate space distribution is then
\begin{align} \label{rhoaperp}
\tilde{\rho}_{s/\bar{s}}({\bf a}_\perp)&=\int d^2{\bf q}_\perp e^{-i{\bf q}_\perp\cdot {\bf a}_\perp}\rho_{s/\bar{s}}({\bf q}_\perp)\nn\\
&= \sum_{\lambda_{s/\bar{s}}}\int \frac{dx_{s/\bar{s}}}{\left(1 - x_{s/\bar{s}}\right)^2} \left\vert \tilde{\psi}_{s/\bar{s}} \Big(x_{s/\bar{s}},\frac{{\bf a}_\perp}{1-x_{s/\bar{s}}},\lambda_{s/\bar{s}}\Big) \right \vert^2.
\end{align}
 Here, ${\bf b}_\perp$ is not the usual LF transverse coordinate variable but related according to 
\mbox{${\bf a}_\perp=(1-x){\bf b}_\perp$}. 

As they are related by a Fourier transform, the strange-antistrange asymmetry in ${\bf b}_\perp$-space is equivalent to the asymmetry of the transverse momentum ${\bf k}_\perp$ distribution.   Since there is no privileged direction for an unpolarized nucleon, one should have a nonvanishing strange-antistrange asymmetry of  the longitudinal momentum distribution if the asymmetry of the transverse momentum distribution is nonzero. A positive $F_1^s(Q^2)$ implies that the $s$ quarks in the nucleon sea are more centralized in coordinate space than the $\bar{s}$ quarks and are therefore more spread out in momentum space. This leads to a negative $s(x)-\bar{s}(x)$ distribution at small-$x$ and a positive one at large-$x$.


\section{\label{FluctModel}The Baryon-meson fluctuation model}

We first evaluate the $s(x)-\bar{s}(x)$ distribution in the nucleon  using the baryon-meson fluctuation model of Ref.~\cite{Brodsky:1996hc}.  
As in Ref.~\cite{Ma:1990hka}, we shall focus on the fluctuation of the proton to the $K^+\Lambda^0$ state, the lightest kaon-hyperon configuration and thus the state with the minimum off-shellness in invariant mass.  In this nonperturbative approach the momentum distribution of the constituents is maximal at minimum off-shellness;  {\it i.e.}, at equal rapidity: $x_i \simeq m^2_{\perp i}/ \sum_j^N m^2_{\perp j}$. Thus the mean LF momentum fraction of each constituent is proportional to its transverse mass:  $m_{\perp i} = \sqrt { \mathbf{k}^2_{\perp i} + m^2_i}$.

Instead of expanding directly in terms of quarks and gluons as in Eq.~\eqref{fockexp}, the expansion in the fluctuation model is performed using a two-level convolution approach~\cite{Ma:1990hka} in which the proton state is expanded as  the valence state plus the baryon-meson state $|\rm BM\rangle$. The component baryon and the meson wave functions are then further expanded into their quark and gluon Fock states.    This LF cluster-decomposition procedure~\cite{Brodsky:1985gs} for the baryon LFWF is similar to the expansion in the meson cloud model~\cite{Sullivan:1971kd,Thomas:1983fh,Burkardt:1991di,Boros:1998qh}. Considering only the fluctuation to  the $|{\rm BM}\rangle=|\Lambda K\rangle$ state, the expansion yields
\begin{align}
|p\rangle&=\int\frac{dx_{\Lambda}d^2{\bf k}_{{\Lambda}\perp}}{2\sqrt{x_{\Lambda}}(2\pi)^3}
\frac{dx_{K}d^2{\bf k}_{K\perp}}{2\sqrt{x_{K}}(2\pi)^3}
16\pi^3\delta(1-x_{\Lambda}-x_{K})\delta^{(2)}({\bf k}_{{\Lambda}\perp}+{\bf k}_{K\perp})
\Psi(x_{\Lambda},{\bf k}_{\Lambda\perp}, x_K, {\bf k}_{K\perp})|\Lambda K\rangle+\cdots,\label{pexp}
\end{align}
where ``$\cdots$'' represents  states other than $|\Lambda K\rangle$ in the expansion, $x_{\Lambda/K}$ is the longitudinal LF momentum fraction carried by the $\Lambda$/$K$,  and ${\bf k}_{\Lambda/K\perp}$ is the intrinsic transverse momentum of the $\Lambda$/$K$. 

The wave function is normalized to the probability of the fluctuation:
\begin{align}
\int\frac{dx_{\Lambda}d^2{\bf k}_{\Lambda\perp}}{16\pi^3}\int\frac{dx_{K}d^2{\bf k}_{K\perp}}{16\pi^3} 16\pi^3 \delta(1-x_\Lambda-x_K)
\delta^{(2)}({\bf k}_{\Lambda\perp}+{\bf k}_{K\perp})
|\Psi(x_{\Lambda},{\bf k}_{\Lambda\perp}, x_K, {\bf k}_{K\perp})|^2= I_s,
\end{align}
where $I_s$ is the intrinsic strange quark number in \eqref{snorm}.

The intrinsic distribution $s(x)$ is then expressed as a convolution of the strange distribution $ q_{s/\Lambda}$ in the $\Lambda$ and the $\Lambda$ distribution $f_{\Lambda/\Lambda K}$ in the baryon-meson state,
\begin{align}
s(x)=\int_x^1\frac{dx_\Lambda}{x_\Lambda} f_{\Lambda/\Lambda K}(x_\Lambda) q_{s/\Lambda}\Big(\frac{x}{x_\Lambda}\Big).
\end{align}
Likewise, the intrinsic distribution $\bar{s}(x)$ is
\begin{align}
\bar{s}(x)=\int_x^1\frac{dx_K}{x_K} f_{K/\Lambda K}(x_K) q_{\bar{s}/K}\Big(\frac{x}{x_K}\Big).
\end{align}
The $\Lambda$ and $K$ distributions in the baryon-meson state are
\begin{align}
f_{\Lambda/\Lambda K}(x_\Lambda)&=\int\frac{d^2{\bf k}_{\Lambda\perp}}{16\pi^3}|\psi_{\Lambda K}(x_\Lambda,{\bf k}_{\Lambda\perp})|^2,\\
f_{K/\Lambda K}(x_K)&=\int\frac{d^2{\bf k}_{K\perp}}{16\pi^3}|\psi_{K\Lambda}(x_K, {\bf k}_{K\perp})|^2,
\end{align}
where
\begin{align}
\psi_{\Lambda K}(x_\Lambda,{\bf k}_{\Lambda\perp})&=\int dx_K d^2{\bf k}_{K\perp}\delta(1-x_\Lambda-x_K)\delta^{(2)}({\bf k}_{\Lambda\perp}+{\bf k}_{K\perp})\Psi(x_{\Lambda},{\bf k}_{\Lambda\perp}, x_K, {\bf k}_{K\perp}),\\
\psi_{K\Lambda}(x_K,{\bf k}_{K\perp})&=\int dx_\Lambda d^2{\bf k}_{\Lambda\perp}\delta(1-x_\Lambda-x_K)\delta^{(2)}({\bf k}_{\Lambda\perp}+{\bf k}_{K\perp})\Psi(x_{\Lambda},{\bf k}_{\Lambda\perp}, x_K, {\bf k}_{K\perp}).
\end{align}
One can observe that
\begin{align}
\psi_{\Lambda K}(x,{\bf k}_\perp)&=\psi_{K\Lambda}(1-x,-{\bf k}_\perp),
\end{align}
which leads to the relation
\begin{align}
f_{\Lambda/\Lambda K}(x)=f_{K/\Lambda K}(1-x).\label{symrel}
\end{align}
The equal numbers of strange and antistrange quarks in the nucleon, {\it i.e.}, Eq.~\eqref{firstmoment}, is satisfied by the sum rules 
\begin{align}
\int_0^1 dx \, q_{s/\Lambda}(x)&=1,\\
\int_0^1 dx \, q_{\bar{s}/K}(x)&=1.
\end{align}
However, the distribution $s(x)-\bar{s}(x)$ remains nontrivial.

We now calculate $F_1^s(Q^2)$. For definitive predictions we adopt the approach used in Ref.~\cite{Vega:2015hti}, in which the $s$ quark wave function is evaluated from the strange quark-scalar diquark configuration $|sD\rangle$ of the $\Lambda$, and the $\bar{s}$ quark is evaluated from the antistrange quark-spectator quark configuration $|\bar{s}q\rangle$ of the $K$. Similar to the expansion~\eqref{pexp}, the $\Lambda$ and $K$ states are expressed as
\begin{align}
|\Lambda\rangle&=\int\frac{dx_s d^2{\bf k}_{s\perp}}{16\pi^3\sqrt{x_s(1-x_s)}}\psi_{sD}(x_s,{\bf k}_{s\perp})|sD\rangle+\cdots,\\
|K\rangle&=\int\frac{dx_{\bar{s}}d^2{\bf k}_{\bar{s}\perp}}{16\pi^3\sqrt{x_{\bar{s}}(1-x_{\bar{s}})}}\psi_{\bar{s}q}(x_{\bar{s}},{\bf k}_{\bar{s}\perp})|\bar{s}q\rangle+\cdots.
\end{align}
Then $F_1^s(Q^2)$ can be expressed in terms of the LFWFs as
\begin{align}
F_1^s(Q^2)&=\mathcal{F}_{s/\Lambda}(Q^2)\mathcal{F}_{\Lambda/p}(Q^2)-\mathcal{F}_{\bar{s}/K}(Q^2)\mathcal{F}_{K/p}(Q^2),
\end{align}
where
\begin{align}
\mathcal{F}_{s/\Lambda}(Q^2)&=\int\frac{dx_s d^2{\bf k}_{s\perp}}{16\pi^3}\psi_{sD}^*(x_s,{\bf k}_{s\perp}+(1-x_s){\bf q}_\perp)\psi_{sD}(x_s,{\bf k}_{s\perp}),\\
\mathcal{F}_{\Lambda/p}(Q^2)&=\int\frac{dx_\Lambda d^2{\bf k}_{\Lambda\perp}}{16\pi^3}\psi_{\Lambda K}^*(x_\Lambda,{\bf k}_{\Lambda\perp}+(1-x_\Lambda){\bf q}_\perp)\psi_{\Lambda K}(x_\Lambda,{\bf k}_{\Lambda\perp}),\\
\mathcal{F}_{\bar{s}/K}(Q^2)&=\int\frac{dx_{\bar{s}} d^2{\bf k}_{\bar{s}\perp}}{16\pi^3}\psi_{\bar{s}q}^*(x_{\bar{s}},{\bf k}_{\bar{s}\perp}+(1-x_{\bar{s}}){\bf q}_\perp)\psi_{\bar{s}q}(x_{\bar{s}},{\bf k}_{\bar{s}\perp}),\\
\mathcal{F}_{K/p}(Q^2)&=\int\frac{dx_K d^2{\bf k}_{K\perp}}{16\pi^3}\psi_{K\Lambda}^*(x_K,{\bf k}_{K\perp}+(1-x_K){\bf q}_\perp)\psi_{K\Lambda}(x_K,{\bf k}_{K\perp}).
\end{align}

For the phenomenological description of the LFWFs, we choose the Brodsky-Huang-Lepage prescription~\cite{Brodsky:1980vj,Brodsky:1981jv} as utilized in Ref.~\cite{Ma:1990hka}, {
\begin{align}
\phi(x,{\bf k}_\perp)&=N\exp\Big[-\frac{1}{8\beta^2}\Big(\frac{{\bf k}_\perp^2}{x(1-x)}+\mathcal{M}_{12}^2\Big)\Big],
\end{align}
with invariant mass
\begin{align}
\mathcal{M}_{12}^2&=\frac{m_1^2}{x}+\frac{m_2^2}{1-x},
\end{align}
where $m_1$ and $m_2$ are the masses of the two components. The $s(x)-\bar{s}(x)$ asymmetry has been calculated with this LFWF in Ref.~\cite{Brodsky:1996hc} and reproduced in Ref.~\cite{Vega:2015hti} with the  parameters $m_q=0.330\,\rm GeV$, $m_s=0.480\,\rm GeV$, $m_D=0.600\,\rm GeV$, and the universal momentum scale $\beta=0.330\,\rm GeV$.   A determination from the data of extended observables indicates $0.24<\beta<0.37\,\rm GeV$~\cite{Zhang:2016qqg}. For the masses of $\Lambda$ and $K$, we use the values given in Ref.~\cite{Patrignani:2016xqp}.  Alternative  LFWFs have been assumed for the study of the $s(x)-\bar{s}(x)$ asymmetry using the same baryon-meson fluctuation picture in Ref.~\cite{Vega:2015hti}.

Taking the fluctuation probability $I_s=1.27\%$ from Ref.~\cite{Cao:1999da}, we calculate $F_1^s(Q^2)$. The results are shown in Fig.~\ref{f1sWF} along with the lattice QCD results~\cite{Sufian:2016pex}.  This result  is consistent with lattice QCD using the original parameters assumed in Ref.~\cite{Brodsky:1996hc}.

\begin{figure}[htp]
\centering
\includegraphics[width=0.6\textwidth]{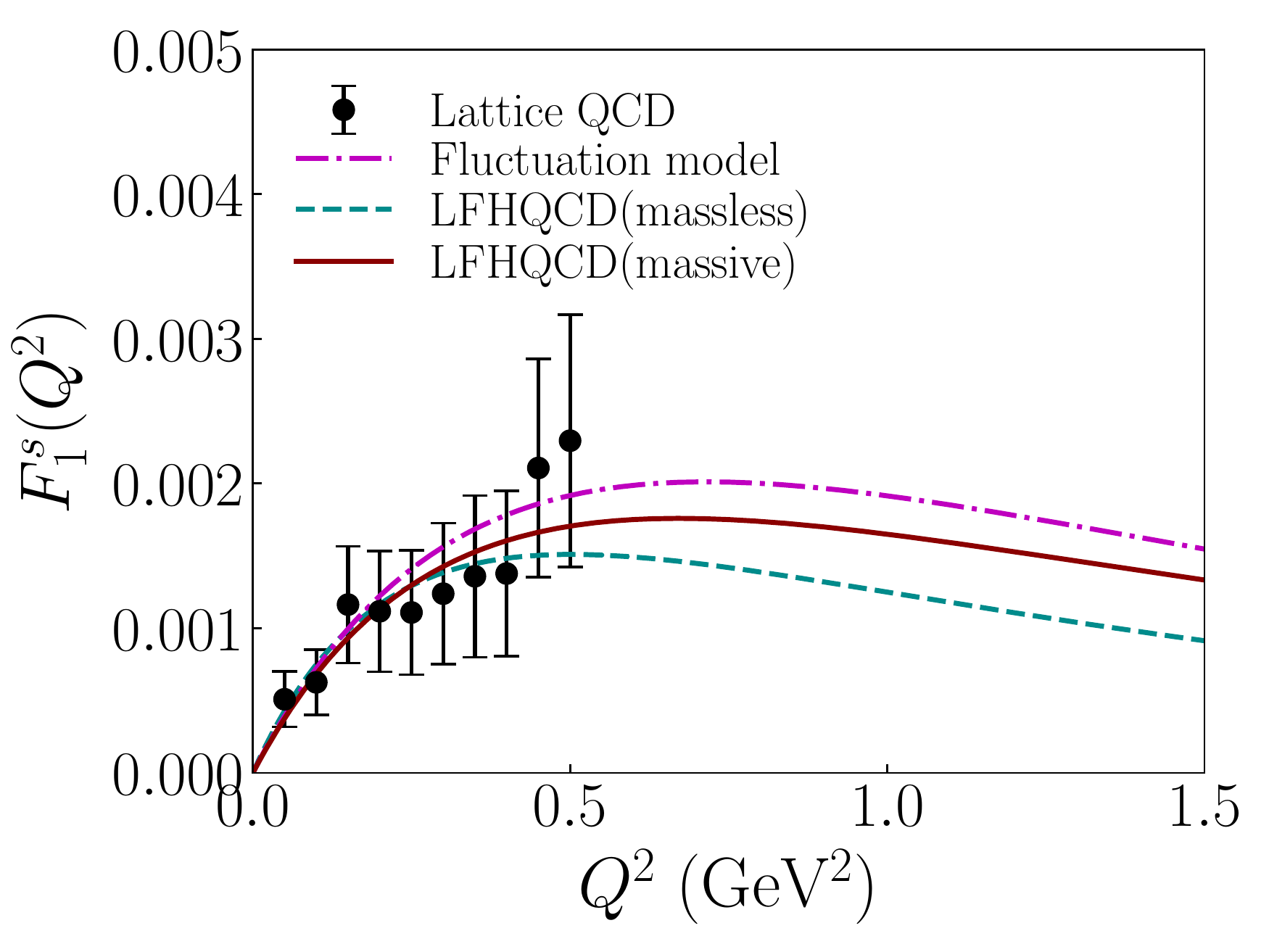}
\caption{Predictions for $F_1^s(Q^2)$ from the fluctuation model, LFHQCD, and  lattice QCD~\cite{Sufian:2016pex,Sufian:2017osl}.  The predictions of the fluctuation model use the  LFWFs from Refs.~\cite{Brodsky:1980vj,Brodsky:1981jv}.}
\la{f1sWF}
\end{figure}

To further constrain the baryon-meson fluctuation model, we will match its predictions to the lattice QCD data by taking $\beta$ and $I_s$ as free parameters. The result is shown in Fig.~\ref{f1sfit}, with $\beta=0.31(11)\,\rm GeV$ and $I_s=1.06(51)\%$. These values are consistent with the original choice in Ref.~\cite{Brodsky:1996hc} and the value determined in Ref.~\cite{Zhang:2016qqg}.

If we take the model parameters determined by the fits, we obtain a model-based phenomenological constraint on the $s(x)-\bar{s}(x)$  distribution  based on the baryon-meson fluctuation approach.  A comparison with global PDF fits is shown in Fig.~\ref{pdfcompare}.  The factorization scale is not specified in this nonperturbative model, so the comparison has been done assuming $\mu=1\,\rm GeV$. The PDF uncertainties are commonly represented in two ways: the Hessian matrix and the Monte Carlo samplings. In Fig.~\ref{pdfcompare}, the uncertainty bands of the global fits are standard deviations calculated from the Hessian matrix for MMHT2014~\cite{Harland-Lang:2014zoa} and JR14~\cite{Jimenez-Delgado:2014twa} and from Monte Carlo replicas for NNPDF3.0~\cite{Ball:2014uwa}. The Monte Carlo replicas for MMHT2014 and JR14 are generated from the Hessian matrix following the procedure described in Ref.~\cite{Hou:2016sho}.

\begin{figure}[htp]
\centering
\includegraphics[width=0.6\textwidth]{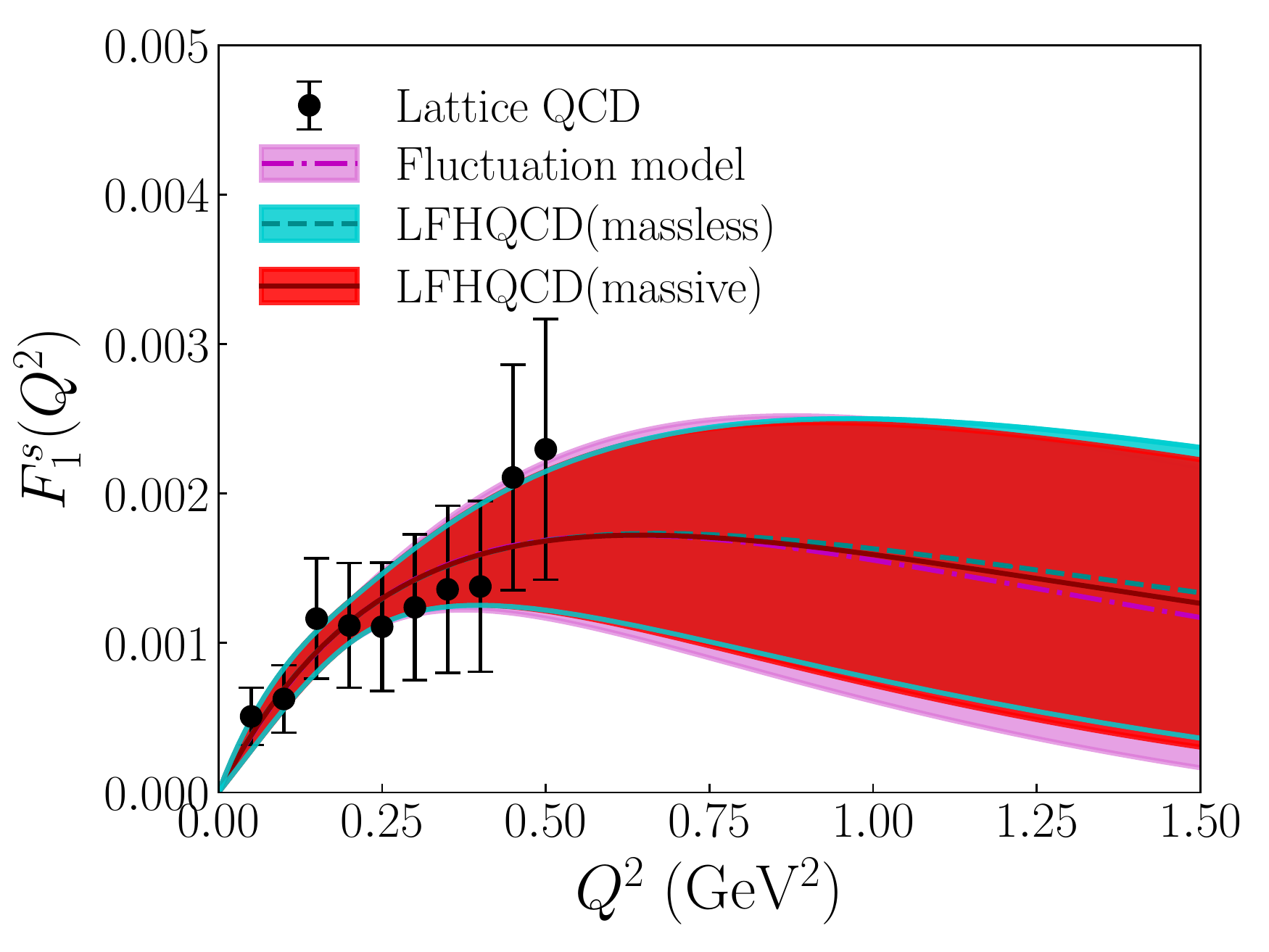}
\caption{Fits to the lattice QCD data of $F_1^s(Q^2)$ using the fluctuation model and LFHQCD.}\label{f1sfit}
\end{figure}


\section{Light-front holographic QCD \label{HologModel}}

\begin{figure}[htp]
\centering
\includegraphics[width=0.55\textwidth]{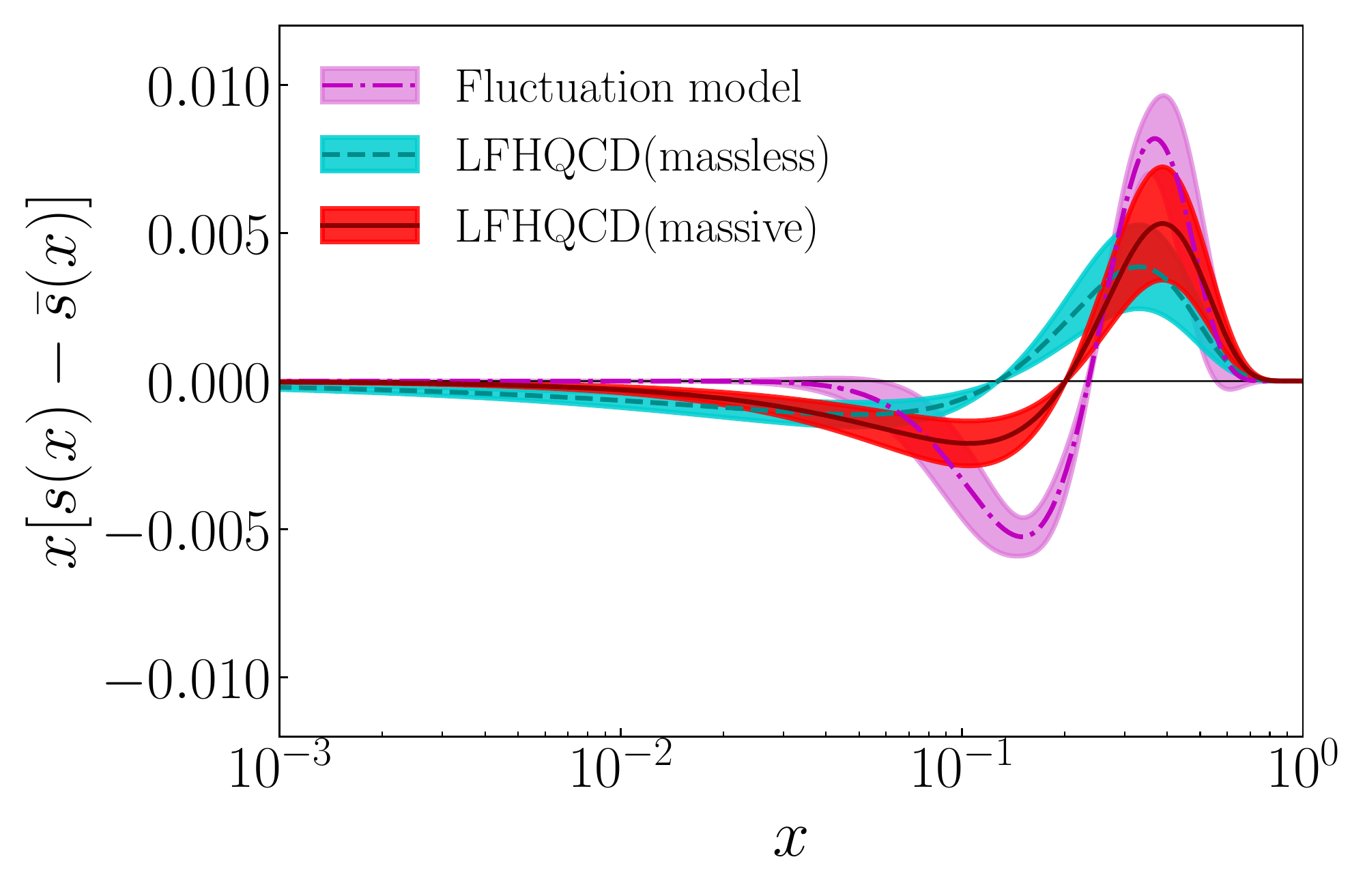}
\includegraphics[width=0.55\textwidth]{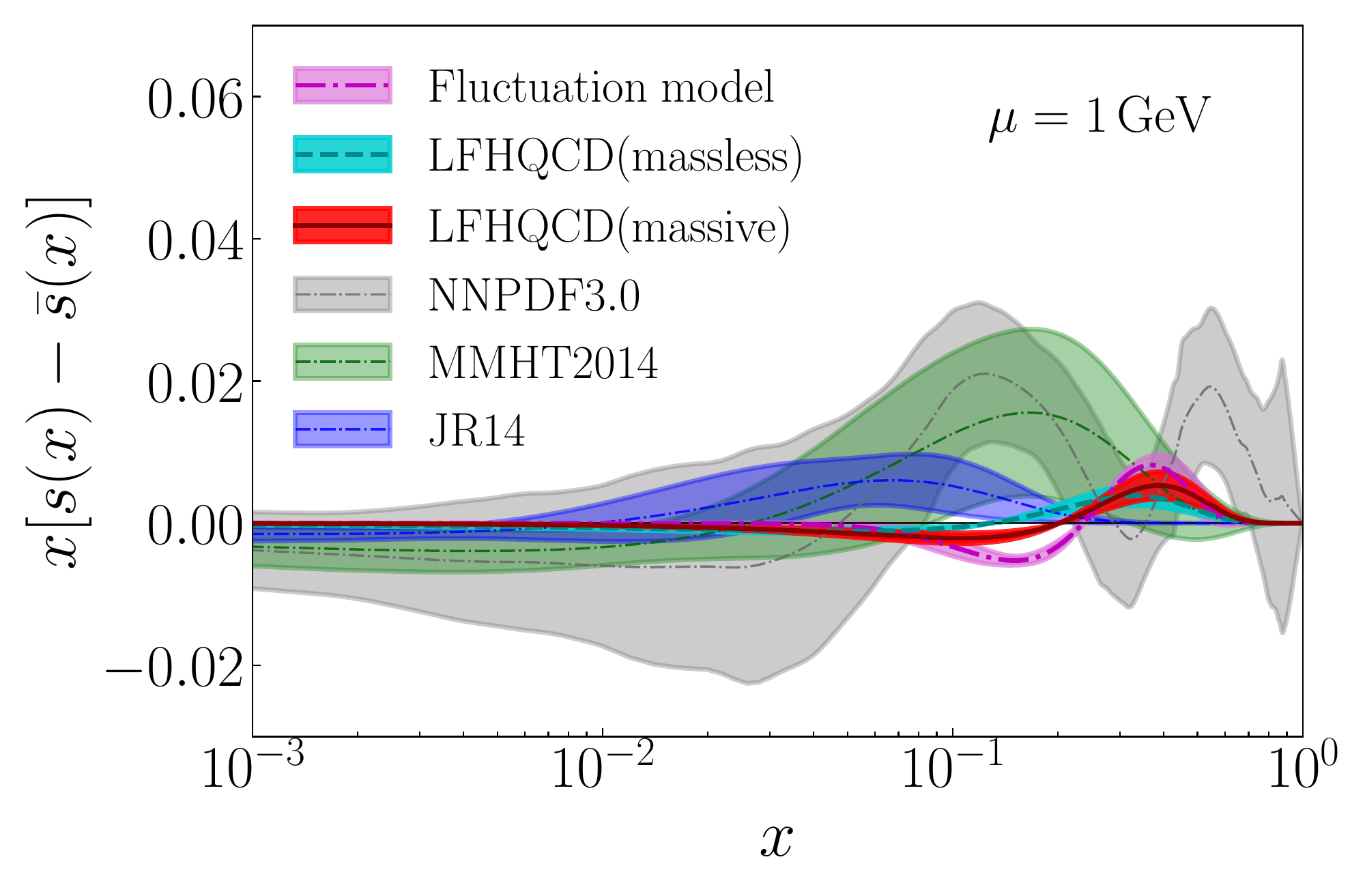}
\includegraphics[width=0.55\textwidth]{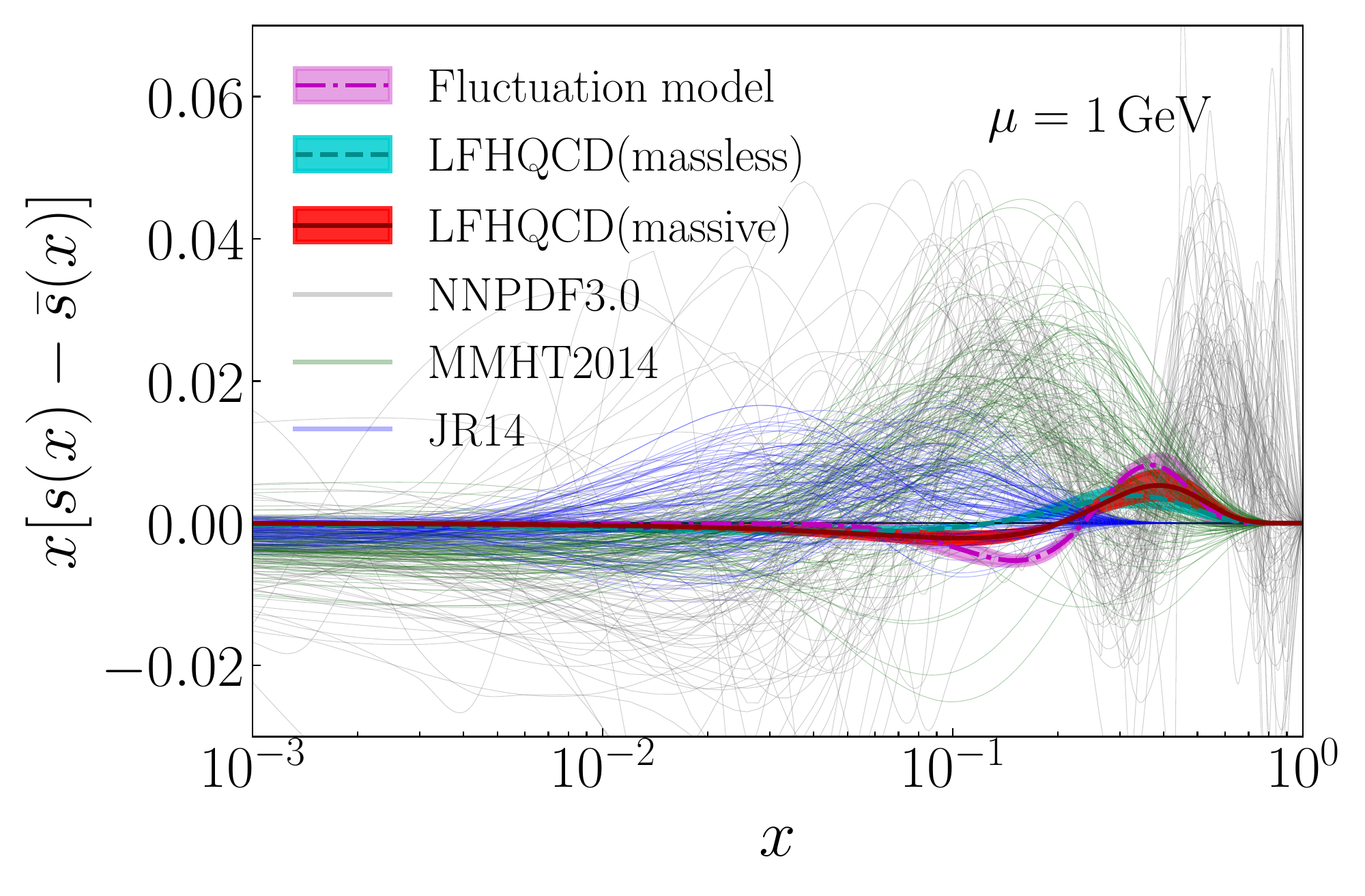}
\caption{Asymmetric strange-antistrange  $x[s(x)-\bar{s}(x)]$ distribution. In the upper panel, the fit results from the fluctuation model and LFHQCD are compared. In the middle panel, the global fits are presented by central curves and standard deviation bands. In the lower panel, the global fits are presented by a hundred Monte Carlo replicas. The global fits are at $\mu=1\,\rm GeV$: NNPDF3.0 (gray)~\cite{Ball:2014uwa},  MMHT2014 (green)~\cite{Harland-Lang:2014zoa}, JR14 (cyan)~\cite{Jimenez-Delgado:2014twa}.}\label{pdfcompare} 
\end{figure}

The EM form factors of nucleons were described in the nonperturbative holographic framework  from the coupling of the $\rho$ to a $q \bar q$ pair in the proton in the limit of massless quarks~\cite{Sufian:2016hwn}. In this section we  calculate  $F_1^s(Q^2)$ and  $s(x)-\bar{s}(x)$  in the  nucleon using the analytic structure of form factors and quark distribution functions in  LFHQCD for bound states of arbitrary twist.   Here, twist refers to the dimension minus spin of the interpolating operator for the hadron state;  it is equal to the number of constituents in a given Fock component in the  LF Fock expansion.  

In LFHQCD~\cite{Brodsky:2014yha}, the EM form factors for a  bound-state hadron with twist-$\tau$ can be expressed as~\cite{deTeramond:2018ecg, Zou:2018eam} 
\begin{align} \label{FFBtau}
F_\tau(t) &= \frac{1}{N_\tau} B\big(\tau-1,  1 - \alpha(t) \big),
\end{align}
where the Euler Beta function is 
\begin{align} \label{intB}
B(u,v)= \int_0^1  dy\, y^{u -1} \, (1-y)^{v -1},
\end{align}
with $B(u,v) =  B(v, u)=\frac{\Gamma(u) \Gamma(v)}{\Gamma(u + v)}$, $N_\tau=\Gamma(\tau-1)\Gamma(1-\alpha(0))/\Gamma(\tau-\alpha(0))$ a normalization factor,  and $\alpha(t)$ is the Regge trajectory of the vector meson which couples to the EM current in the $t$-channel  exchange.

The Beta function structure of the EM form factors~\eqref{FFBtau}, which follows from the gauge/gravity structure in LFHQCD, was obtained in the pre-QCD era by Ademollo and Del~Giudice~\cite{Ademollo:1969wd} and independently by  Landshoff and Polkinghorne~\cite{Landshoff:1970ce}. Their derivations were based on the Veneziano model~\cite{Veneziano:1968yb}, which is an incorporation of the concept of duality~\cite{Dolen:1967jr} in a pole model. For hadronic four-point functions, it leads to a representation of the scattering amplitude by Euler Beta functions. Extending these considerations to current induced interactions, a structure like~\eqref{FFBtau} was derived in Refs.~\cite{Ademollo:1969wd, Landshoff:1970ce,Bender:1970ew}. However, the variable $\tau$ in the duality based derivations is a free parameter and the Regge trajectory is a phenomenological input. In contradistinction, LFHQCD provides a clear physical meaning  of $\tau$, the twist of a given Fock component of the hadron, and also incorporates the Regge trajectory from the vector-meson  (VM) spectrum by solving the semiclassical LF QCD Hamiltonian eigenvalue problem.

For linear Regge trajectories 
\begin{align} \label{CF}
\alpha(t) = \alpha(0) + \alpha' t,
\end{align}
Eq. \eqref{FFBtau}  incorporates  the hard-scattering counting rules at large $t$~\cite{Brodsky:1973kr, Matveev:ra}. Indeed, for fixed $u$ and large $v$ we have $B(u,v) \sim \Gamma(u) v^{-u}$, and therefore the first argument in the Euler Beta function  determines the scaling behavior of \eqref{FFBtau}
\begin{align} \label{FFas}
\lim_{Q^2 \to \infty} F_\tau(Q^2) = \Gamma(\tau - 1)  \left(\frac{1}{\alpha 'Q^2}\right)^{\tau-1} ,
\end{align}
at large $Q^2 = -t$. The second argument in \eqref{FFBtau} determines the timelike pole structure of the form factor; the analytic structure of (\ref{FFBtau}) thus leads to a nontrivial connection with the hadron spectrum.  In fact, using the expansion of the Gamma function 
\begin{align}
\Gamma(N + z) = (N - 1 + z) (N - 2 + z) \cdots (1 + z) \Gamma(1 + z),
\end{align} 
for integer twist $N = \tau$, with $N$ the number of constituents  for a given Fock component, we find 
\begin{align} \label{FtauM}
F_{\tau}( Q^2) = \frac{1}{\left(1 + \frac{Q^2}{M^2_{n=0}}\right) \left(1 + \frac{Q^2}{M^2_{n=1}} \right) \cdots \left(1 + \frac{Q^2}{M^2_{n =\tau - 2}} \right)},
\end{align}
which is expressed as a product of $\tau -1$ poles located at 
\begin{align}  \label{M2RT}
 - Q^2 = M^2_n = \frac{1}{\alpha'}\bigg(n + 1 - \alpha(0)\bigg).
\end{align}
The form factor \eqref{FtauM} thus generates the radial excitation spectrum of the exchanged particles in the $t$-channel, while keeping  the structural form found previously in the limit of zero quark masses~\cite{Brodsky:2014yha}.

For the lowest radial excitation the VM spectrum in LFHQCD is given by~\cite{Brodsky:2014yha, Brodsky:2016yod} (Appendix~\ref{VMRT})
\begin{align} \label{M2VM} 
M^2 = 4 \lambda\left(J -\frac{1}{2}\right) + \Delta M^2,
\end{align}
where the squared mass shift $\Delta M^2$ incorporates the effect from finite light quark masses.  The quantity $\lambda = \kappa^2$ is the emergent mass scale,  the  only  dimensional quantity appearing in LFHQCD for massless quarks~\cite{Brodsky:2014yha}. Its value determined from the best fit to all radial and orbital excitations of the light mesons and baryons is $\kappa = \sqrt \lambda = 0.523 \pm 0.024\,\rm GeV$~\cite{Brodsky:2016yod}.

\begin{figure}[htp] 
\begin{center} 
\includegraphics[width=8.0cm]{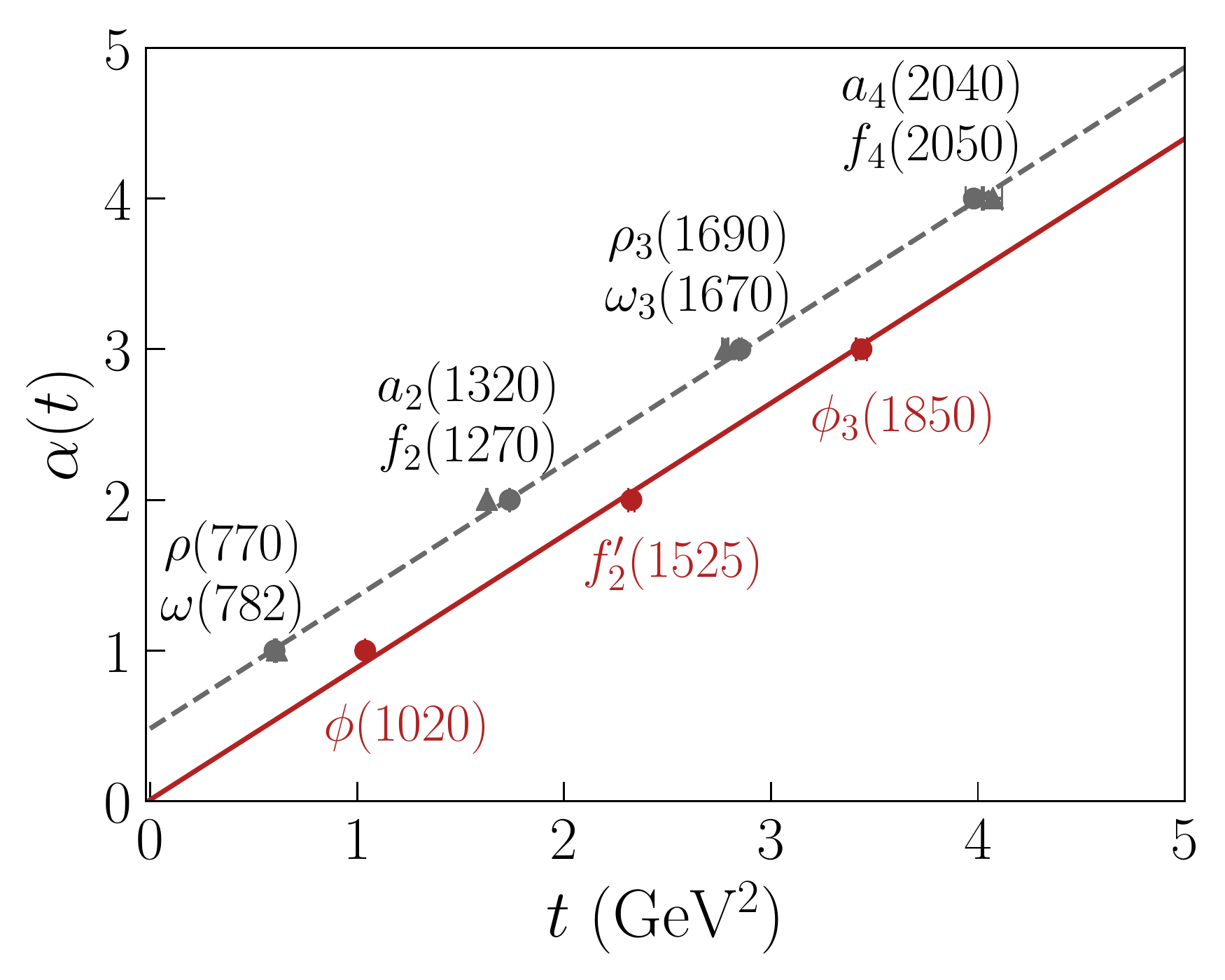}
\caption{Chew-Frautschi plot for the leading $\rho$  and $\omega$ (gray dashed) and $\phi$ (red continuous) trajectories in LFHQCD. At values $t=M^2$ where $\alpha(t)$ is an integer, there is a hadron with mass squared $M^2$ and spin $J= \alpha(M^2)$. The $\rho$ and $\omega$ intercepts are fixed by the pion mass  from the relation $\Delta M_\rho^2 = \Delta M_\omega^2 = M^2_{\pi^\pm}$ and the mass scale $\lambda$ is fixed by the best fit to the slopes of both trajectories: This fixes the intercept of the $\phi$ trajectory. We find $\sqrt{\lambda} = 0.534  \, {\rm GeV}$, $\alpha_\rho(0) = \alpha_\omega(0) = \frac{1}{2}  - \frac{\Delta M_\pi^2}{4 \lambda} = 0.483$  and $\alpha_\phi(0) =  0.01$. Solid triangles represent the $\omega$ trajectory. The data is from Ref.~\cite{Patrignani:2016xqp}.}
\label{C-F}
\end{center}
\end{figure}

There is no need to introduce additional procedures to include quark masses when using the structural form \eqref{FFBtau} to describe  form factors, since the effect of quark masses only amounts to a shift of the Regge intercept. For example, for the $\rho, a$ vector mesons we obtain from  Eq.~\eqref{M2VM} the leading Regge trajectory 
\begin{align} \label{alpharho}
\alpha_\rho(t) = \frac{1}{2}+ \frac{t}{4 \lambda} -  \frac{ \Delta M_\rho^2}{4 \lambda},
\end{align}
with  slope $\alpha' = \frac{1}{4 \lambda}$ and intercept $\alpha_\rho(0) = \frac{1}{2}  - \frac{\Delta M_\rho^2}{4 \lambda}$, which differs from the conformal limit $\half$ by the mass shift $\frac{\Delta M_\rho^2}{4\lambda}$ from quark masses. Likewise, the $\omega, f$ trajectory is
\begin{align} \label{alphaom}
\alpha_\omega(t) = \frac{1}{2}+ \frac{t}{4 \lambda} -  \frac{ \Delta M_\omega^2}{4 \lambda},
\end{align}
with the same slope $\alpha' = \frac{1}{4 \lambda}$ and similar intercept $\alpha_\omega(0) = \frac{1}{2}  - \frac{\Delta M_\omega^2}{4 \lambda}$. We show in Fig.~\ref{C-F}  the Chew-Frautschi plot for the leading  $\rho-a$ and $\omega-f$ trajectories.

The spectrum of the exchanged particles in the $t$-channel  follows from \eqref{M2RT} for the leading VM trajectory \eqref{alpharho}. We find
\begin{align}
-Q^2 = M^2 = 4 \lambda\left(n + \frac{1}{2}\right) + \Delta M_\rho^2,
\end{align}
which is precisely the  spectrum of the $\rho$ and its radial excitations~\cite{Brodsky:2014yha} (Appendix \ref{VMRT}). In this case the shift in the intercept is rather small since ${\Delta M_\rho^2} = \Delta M_\omega^2 = M^2_{\pi^\pm}$ and $\frac{M^2_\pi}{4 \lambda} \simeq 0.02$.

\subsection{Strange quark form factor \label{strangeFF}}

In contrast to the two-step convolution expansion of the fluctuation model, $F_1^s(Q^2)$ and $s(x)-\bar{s}(x)$ from LFHQCD can be obtained directly from higher-twist terms in the Fock state expansion by matching to the quark degrees of freedom. To this end, let us recall that for the up and down quark form factors  the $\rho$-trajectory is relevant because it dominantly couples to $u \bar u$ and $d \bar d$  quark  currents in the proton~\cite{Sufian:2016hwn}. Likewise, we compute  $F_1^s(Q^2)$  in the holographic framework by considering the Regge trajectory of the $\phi$ meson, which is nearly a pure $s \bar s$ state~\cite{Amsler:PDG}, and therefore couples dominantly to the $s \bar s$ sea current in the nucleon.

To determine the slope and intercept of the $\phi$ trajectory, 
\begin{align} \label{alphaphi}
\alpha_\phi(t) = \frac{1}{2}+ \frac{t}{4 \lambda} - \frac{\Delta M_\phi^2}{4 \lambda},
\end{align}
we fix the $\rho$ intercept from the pion mass and find the best value for the universal Regge slope  from the simultaneous fit of the $\rho$ and $\phi$ trajectories;  this procedure determines the $\phi$ intercept and the universal slope $\alpha' = \frac{1}{4 \lambda}$. We obtain $\sqrt{\lambda} = 0.534 \, {\rm GeV}$ and $\alpha_\phi(0) =  0.01$, or equivalently $\Delta  M_\phi^2 = 1.96 \, \lambda$.  The $\phi-f'$ trajectory is shown in Fig.~\ref{C-F}.  One can also compute the intercept in LFHQCD with effective quark masses, see Appendix~\ref{VMRT}, the value is $\alpha_\phi(0)=0.00 \pm 0.04$. The value of $\Delta M_\phi^2$  is significantly larger than $\Delta M_\rho^2$ due to the presence of the more massive strange quarks in the $\phi$ meson.

Since the light-front holographic framework is inherently relativistic, the LFWF for a state with twist-$\tau$ automatically incorporates Fock state components with two different orbital angular momenta $L^z$ and $L^z+1$, in analogy to the upper and lower components of a Dirac 4-component spinor.   For example, the valence quark distributions of a nucleon correspond to a leading twist-$3$ effective LFWF with orbital angular momentum $L^z=0$, plus a twist-$4$ term corresponding to a three-quark effective LFWF with $L^z=1$.  Note that Fock states with both $L^z $ and $L^z+ 1$ are needed in order that a baryon can have a nonzero Pauli form factor and a nonzero anomalous magnetic moment~\cite{Brodsky:1980zm}.

The five-quark state $|uuds\bar{s}\rangle$ is the lowest Fock state which contains strangeness. Therefore, the leading contributions to the strange form factor are terms with twist-$5$ and twist-$6$.  Using the constraint $F_1^s(0)=0$ from the sum rule~\eqref{firstmoment}, the analytic structure of $F_1^s(Q^2)$ is uniquely determined by the holographic structure up to twist-6:
\begin{align} \label{F1s}
F_1^s(Q^2)= (1 - \eta)N_s\left[F^\phi_{\tau=5}(Q^2)-F^\phi_{\tau=6}(Q^2)\right] + \eta N_s\left[F^\omega_{\tau=5}(Q^2)-F^\omega_{\tau=6}(Q^2)\right],
\end{align}
where we have allowed for a small $\phi-\omega$ mixing $\eta$ in the strange form factor~\cite{Jaffe:1989mj}. $N_s$ is a normalization factor and $F^{\omega, \phi}_\tau(Q^2)$ is the twist-$\tau$ form factor \eqref{FFBtau} with Regge trajectory $\alpha_{\omega,\phi}(t)$ given by \eqref{alphaom} and \eqref{alphaphi} respectively. The form factor can also be expressed as a product of $\tau -1$ poles located at $t= - Q^2 = 4 \lambda \left(n+\half \right) +  \Delta M_\omega^2$ and $t= - Q^2 = 4 \lambda \left(n+\half \right) +  \Delta M_\phi^2$, $n=0, 1, 2 \cdots {\tau-2}$. One thus obtains in this case the form factor poles at the mass of the $\omega$ and $\phi$ vector meson and its radial excitations.

\begin{figure}[htp]
\centering
\includegraphics[width=0.55\textwidth]{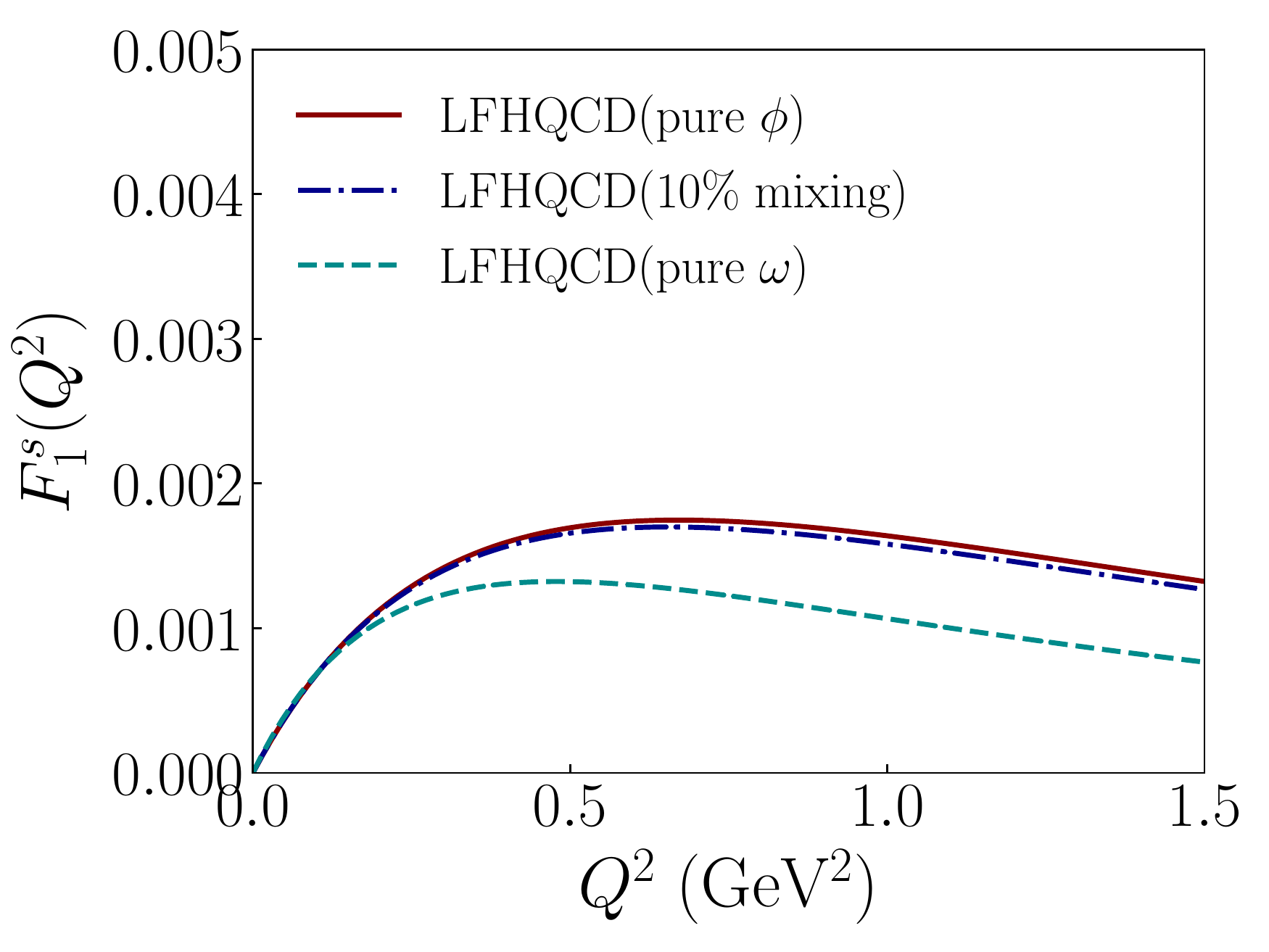}
\includegraphics[width=0.55\textwidth]{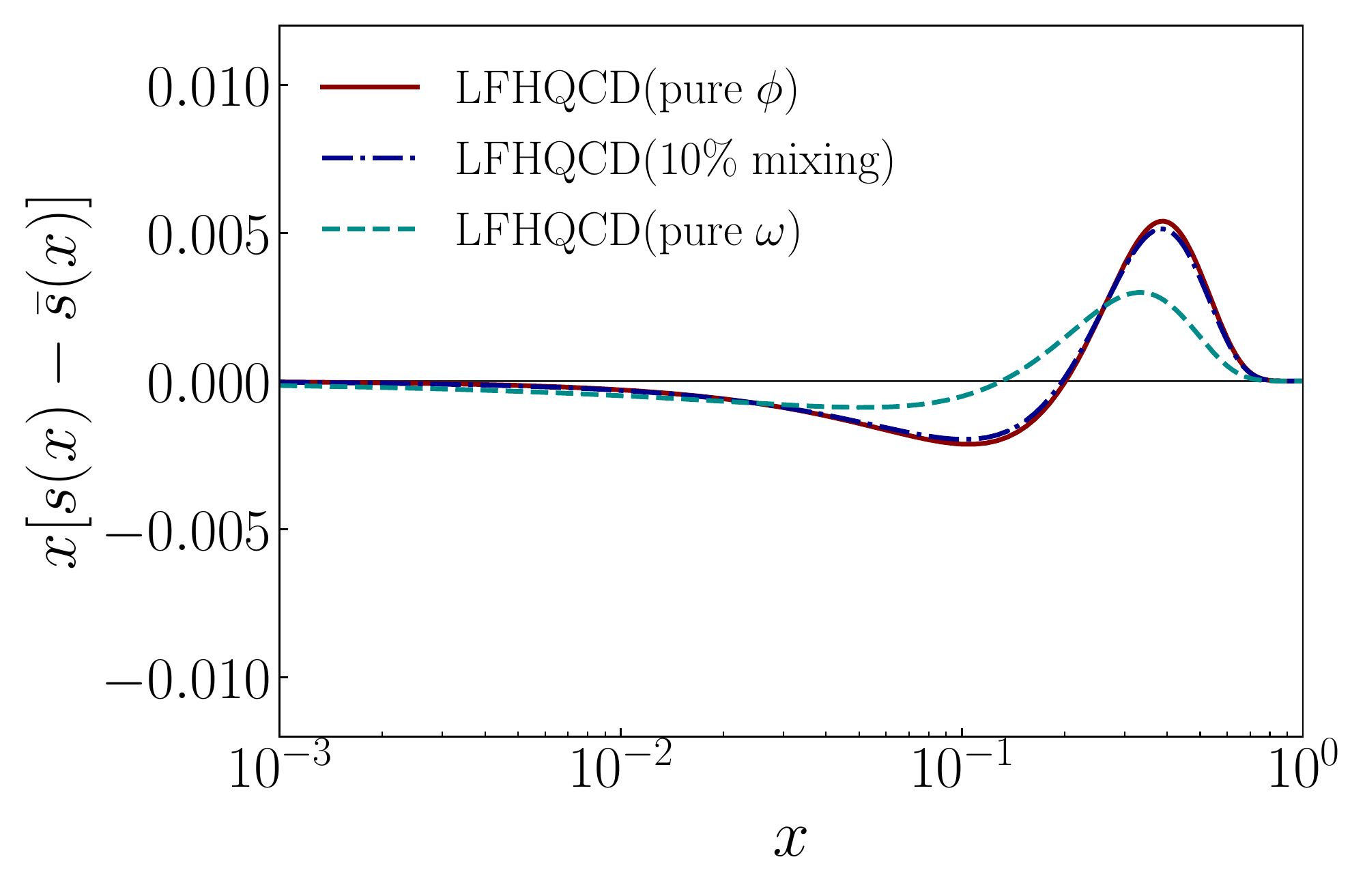}
\caption{Effect of $\phi-\omega$ mixing in $F_1^s(Q^2)$ and the $s(x) - \bar s(x)$ asymmetry. The effect of the mixing is negligible even for 10\% mixing, {\it i.e.}, for $\eta=0.1$.}
\label{phirhomix} 
\end{figure}

To illustrate the effect of the $\phi-\omega$ mixing we show in Fig.~\ref{phirhomix} the effect of a 10 \% mixing in $F_1^s(Q^2)$. The effect of the small mixing turns out to be negligible for $F_1^s(Q^2)$. We also show in Fig.~\ref{phirhomix} the chiral limit for massless quarks. Since the quark mass effect is very small in the $\omega$ trajectory, this chiral limit corresponds to a pure $\omega$ trajectory.

Note that the normalization factor $N_s$ in \eqref{F1s} is not the intrinsic strange/antistrange quark number $I_s$, since the strange and antistrange distributions can both have twist-$5$ and twist-$6$ contributions.   However, the shape of $F_1^s(Q^2)$ is completely determined from the structure of LFHQCD. The result is shown in Fig.~\ref{f1sWF}, together with predictions from the fluctuation model and lattice QCD.  The value of $\sqrt \lambda = 0.534$ GeV and the mass shift $\Delta M_\phi^2 = 1.96 \,\lambda$ are obtained from the $\phi$ trajectory depicted in Fig.~\ref{C-F}. The value of $N_s= 0. 047$  in Fig.~\ref{f1sWF} is determined by a best fit to lattice QCD predictions. As in the case of the fluctuation model, we also fit the lattice QCD data, taking $\sqrt\lambda$ and $N_s$ as free parameters. The result is shown in Fig.~\ref{f1sfit} with parameter values $\sqrt\lambda = 0.52(17)\,\rm GeV$ and $N_s=0.046(17)$. This value of $\sqrt\lambda$ agrees with  that determined from the Regge trajectory. The conformal limit results, $\Delta M^2=0$, are also shown in  the figures for comparison. The strange form factor \eqref{F1s} has the large-$Q^2$ behavior $Q^8 F_1^s(Q^2) \to Const$, with $Const =  1680 \, N_s \, \lambda^4 \simeq 0.5 \,{\rm GeV}^8$, consistent with the scaling predicted from the hard-scattering  counting rules~\cite{Brodsky:1973kr, Matveev:ra}.

\subsection{Strange quark distribution functions}

To describe the quark distribution functions in the holographic formalism it is convenient to express the Beta function \eqref{intB} in a reparametrization invariant form
\begin{align} \label{Bw}
B(u,v)= \int_0^1  dx \, w'(x) \, w(x)^{u -1} \left(1- w(x)\right)^{v -1},
\end{align}
provided that $w(x)$ satisfies the constraints~\cite{deTeramond:2018ecg}
\begin{align}
w(0) = 0, \quad \quad w(1) = 1, \quad  \quad w'(x) \ge 0.
\end{align}
Therefore, using \eqref{Bw} and the Regge trajectory, \eqref{alpharho}, \eqref{alphaom} or \eqref{alphaphi}, the EM form factor \eqref{FFBtau} for twist-$\tau$ can be written in the invariant form
\begin{align} \label{FFw}
F_\tau(t) = \frac{1}{N_\tau}  \int_0^1 dx \, w'(x) w(x)^{- \frac{t}{4 \lambda} - \half} \big[1- w(x)\big]^{\tau-2} e^{- \frac{\Delta M^2}{4 \lambda} \log\left(\frac{1}{w(x)}\right)}.
\end{align}

The EM form factor  can also be expressed by the exclusive-inclusive connection as the integrated expression of the $t$-evolved PDF, namely, the generalized parton distribution (GPD) at zero skewness, $H^q_\tau(x, t) \equiv H^q_\tau(x ,\xi=0,t)$,
\begin{eqnarray} \label{Fqtau}
F^q_\tau(t) &=& \int_0^1 dx  \left( H^q_\tau(x,t) - H^{\bar q}_\tau(x,t) \right) \nn \\
&=&\int_0^1 dx \, q_\tau(x) \exp[t f(x)], 
\end{eqnarray}
where $f(x)$ is the profile function and $q_\tau(x)$ is the collinear PDF of twist-$\tau$. Comparing \eqref{Fqtau} with the holographic expression \eqref{FFw} we find that both functions, $f(x)$ and $q_\tau(x)$, are determined in terms of the  reparametrization function of the Beta function, $w(x)$,  by 
\begin{align} \label{fw}
f(x)&=\frac{1}{4\lambda}\log\Big(\frac{1}{w(x)}\Big),\\ \label{qw}
q_\tau(x)&=\frac{1}{N_\tau}[1-w(x)]^{\tau-2}w(x)^{-\frac{1}{2}}w'(x) \, e^{- \frac{\Delta M^2}{4 \lambda} \log\left(\frac{1}{w(x)}\right)},
\end{align}
where $q_\tau(x)$ is normalized by $\int_0^1 dx \, q_\tau(x) = 1$.  In the conformal limit where the quark masses vanish, $\Delta M^2 \to 0$, we recover the results given in Ref.~\cite{deTeramond:2018ecg}.

The specific function $w(x)$, taken from Ref.~\cite{deTeramond:2018ecg},  is effectively determined by  Regge behavior at small-$x$ and the local power-law counting rule at $x \to 1$. At $x \to 0$,  $w(x)$ scales as $w(x) \sim x$  to recover Regge behavior~\cite{Goeke:2001tz}. At $x \to 1$  the additional constraints
\begin{align} \label{wp1}
w'(1) = 0 \quad \quad {\rm and} \quad \quad w''(1) \ne 0,
\end{align} 
yield the  Drell-Yan counting rule $q_\tau(x)  \sim (1-x)^{2 \tau - 3}$ at large $x$~\cite{Drell:1969km}. Since $w(1) = 1$,  it follows that $\log\left(\frac{1}{w(x)}\right) \to 0$ in the limit  $x \to 1$, 
which implies that the local counting rules at large-$x$ are unmodified by the introduction of quark masses in the holographic structural framework. However, the squared mass shift induced by finite quark masses does modify the small-$x$ behavior by a factor $x^{\Delta M^2 / 4\lambda}$, therefore softening the  Regge behavior of the PDFs at small-$x$
\begin{align}
q_\tau(x)  \sim x^{-\alpha(0)} \sim x^{- \frac{1}{2} +  \frac{\Delta M^2}{ 4 \lambda}},
\end{align} 
since $w(x)$ in \eqref{qw} scales as $w(x) \sim x$ at small-$x$. Since $\Delta M^2$ is considerably larger for strange quarks than for the up and down quarks, the predicted behavior of the strange sea distributions is less singular at $x \to 0$ than the nonstrange light quarks. 

It has been noted in the pre-QCD era that the behavior of parton distributions near $x \to 0$ is governed by the Regge  intercept~\cite{Landshoff:1970ce}. This is again in agreement with LFHQCD even including the finite quark mass correction.  The $t$-dependence of GPDs, instead, is not influenced by the introduction of quark masses, since the Regge slope is universal for light hadrons ~\cite{Brodsky:2016yod}.

The expression for the strange-antistrange PDF asymmetry $s(x) - \bar{s}(x)$ corresponding to \eqref{F1s} is
\begin{align} \label{s-sb}
s(x) - \bar s(x) =  (1- \eta) N_s \left[q^\phi_{\tau=5}(x) - q^\phi_{\tau=6}(x)\right] +  \eta N_s \left[q^\omega_{\tau=5}(x) - q^\omega_{\tau=6}(x)\right] ,
\end{align}
with $q^{\omega, \phi}_\tau(x)$ given by \eqref{qw} for $\Delta M^2_\omega$ and $\Delta M^2_\phi$ respectively.
For the universal reparametrization function $w(x)$ we use the form in Ref. \cite{deTeramond:2018ecg},
\begin{align} \label{wx}
w(x) = x^{1-x} e^{-a(1-x)^2},
\end{align}
with $a=0.531$ determined from the first moment of proton valence quark distributions. The effect of the $\phi-\omega$ mixing for the $s(x) - \bar s(x)$ asymmetry also turns out to be negligible for a mixing of the order of 10\% and will be neglected.

The PDF predictions for the asymmetry $s(x) - \bar s(x)$ are shown in Fig.~\ref{pdfcompare} and compared with the fluctuation model and global fits for  $N_s=0.046(17)$ and $\sqrt{\lambda} = 0.52(17)\,\rm GeV$ obtained from the lattice form factor results. The actual computations are carried out with the universal function $w(x)$  given by~\eqref{wx}. In contrast to the baryon-meson fluctuation model, which has the  small-$x$ behavior $s(x) - \bar s(x) \to 0$, the holographic model has the Regge behavior  $s(x) - \bar s(x) \simeq -0.044 x^{-0.01}$ in the limit $x \to 0$. This can be compared with the global data fit results, shown in Fig.~\ref{pdfcompare} at the initial scale $\mu = 1\,\rm GeV$.

The sign and the magnitude of $\langle S_-\rangle$, Eq. \eqref{Sm}, play a significant role in understanding the NuTeV anomaly~\cite{Zeller:2001hh,Kretzer:2003wy,Ding:2004ht,Alwall:2004rd,Ding:2004dv,Wakamatsu:2004pd}; namely, that the Weinberg angle $\theta_W$ extracted from deep inelastic neutrino/antineutrino scatterings by NuTeV deviates by about $3\sigma$ from the standard model value $\sin^2\theta_W=0.23129(5)$~\cite{Patrignani:2016xqp}.  A positive $\langle S_-\rangle $ will reduce the NuTeV anomaly, whereas  a negative $\langle S_-\rangle$ will increase it~\cite{Kretzer:2003wy,Alwall:2004rd,Catani:2004nc}.  Assuming a single source for the NuTeV anomaly, $\langle S_- \rangle\sim 0.005$ is required.

From our analysis, the lattice QCD result favors a positive $\langle S_-\rangle $.  However, the fits with the fluctuation model and LFHQCD yield $\langle S_-\rangle=0.0011(4)$, which is not sufficient to solely explain the NuTeV anomaly;  other sources are needed.  Although the value for $\langle S_-\rangle $ is model dependent, we emphasize that more precise determinations of $F_1^s(Q^2)$ from first-principle lattice QCD calculations and/or future experiments will provide important constraints on the strange-antistrange asymmetry.

\subsection{Separation of strange and antistrange asymmetric quark distributions}

Light-front holographic QCD predicts the structural behavior of the strange asymmetry \eqref{s-sb} up to twist-6, but it does not directly predict the individual distributions $s(x)$ and $\bar s(x)$ which together determine the intrinsic strange contribution to the quark sea in the nucleon
\begin{align} \label{intsA}
\int dx \, s(x) = \int dx \, \bar s(x) = I_s.
\end{align}
We will show, however, how one can uniquely determine  the minimum strange probability $I_s$ in the proton  and then give constraints on the separate $s(x)$ and $\bar{s}(x)$ distributions.

We expand the longitudinal quark distributions $s(x)$ and $\bar s(x)$ into their twist-5 and twist-6 components
\begin{eqnarray}
s(x) &=& \alpha \, q_{\tau= 5}(x) +  \beta \,q_{\tau= 6}(x) ,\\
\bar s(x) &=& \gamma \, q_{\tau=5}(x) +  \delta \, q_{\tau =6}(x),
\end{eqnarray}
corresponding to $L^z = 0$ and $L^z = 1$, respectively. Comparing with \eqref{s-sb} and using the sum rule \eqref{firstmoment}, we find
\begin{eqnarray}
\alpha + \beta &=& I_s ,\\
\gamma + \delta &=& I_s ,\\
\alpha - \gamma &=& N_s , \\
\delta - \beta &=& N_s,
\end{eqnarray}
with the general solution
\begin{eqnarray}
\beta &=& I_s - \alpha ,\label{eq63} \\
\gamma &=&  \alpha - N_s,  \\
\delta &=& I_s - \alpha  + N_s  .  
\end{eqnarray}

We can thus write
\begin{eqnarray}
s(x) &=&  \alpha \, q_{\tau= 5}(x) + (I_s - \alpha) \, q_{\tau= 6}(x) ,\\
\bar s(x) &=& (\alpha - N_s) \, q_{\tau=5}(x) +  (I_s - \alpha + N_s) \, q_{\tau =6}(x),
\end{eqnarray}
with $\alpha$ an arbitrary parameter constrained by the conditions $s(x) \ge 0$ and $\bar s(x) \ge 0$. Since the twist-5 term dominates at large-$x$ we require $\alpha \ge 0$ and $\gamma \ge 0$. For positive $N_s$, the positivity constraints lead to $\alpha \ge N_s$. At small-$x$ we have the behavior 
\begin{align}
\lim_{x \to 0} \frac{q_{\tau = 5}(x)}{q_{\tau = 6}(x)} = \frac{N_{\tau=6}}{N_{\tau=5}} \equiv R,
\end{align}
with $N_{\tau}$ defined in \eqref{FFBtau}. In the conformal limit, $\Delta M^2 = 0$, we have $R=\frac{8}{9}$. Incorporating quark masses, $\Delta M_\phi^2 = 1.96 \, \lambda$, we have $R=0.80$. This small-$x$ behavior leads to the condition $I_s \geq (1-R) \alpha$ from Eq. \eqref{eq63}. Together with  $\alpha \ge N_s$ we have the condition 
\begin{align}
N_s  \le \alpha  \le \frac{1}{1-R} I_s. \label{posi}
\end{align}
Because the ratio $q_{\tau= 5}(x)/q_{\tau= 6}(x)$  is monotonically increasing, the condition  \eqref{posi} ensures $s(x) \ge 0$ and $\bar s(x) \ge 0$ over the full range of $x$.

\begin{figure}[htp] 
\begin{center} 
\includegraphics[width=8.56cm]{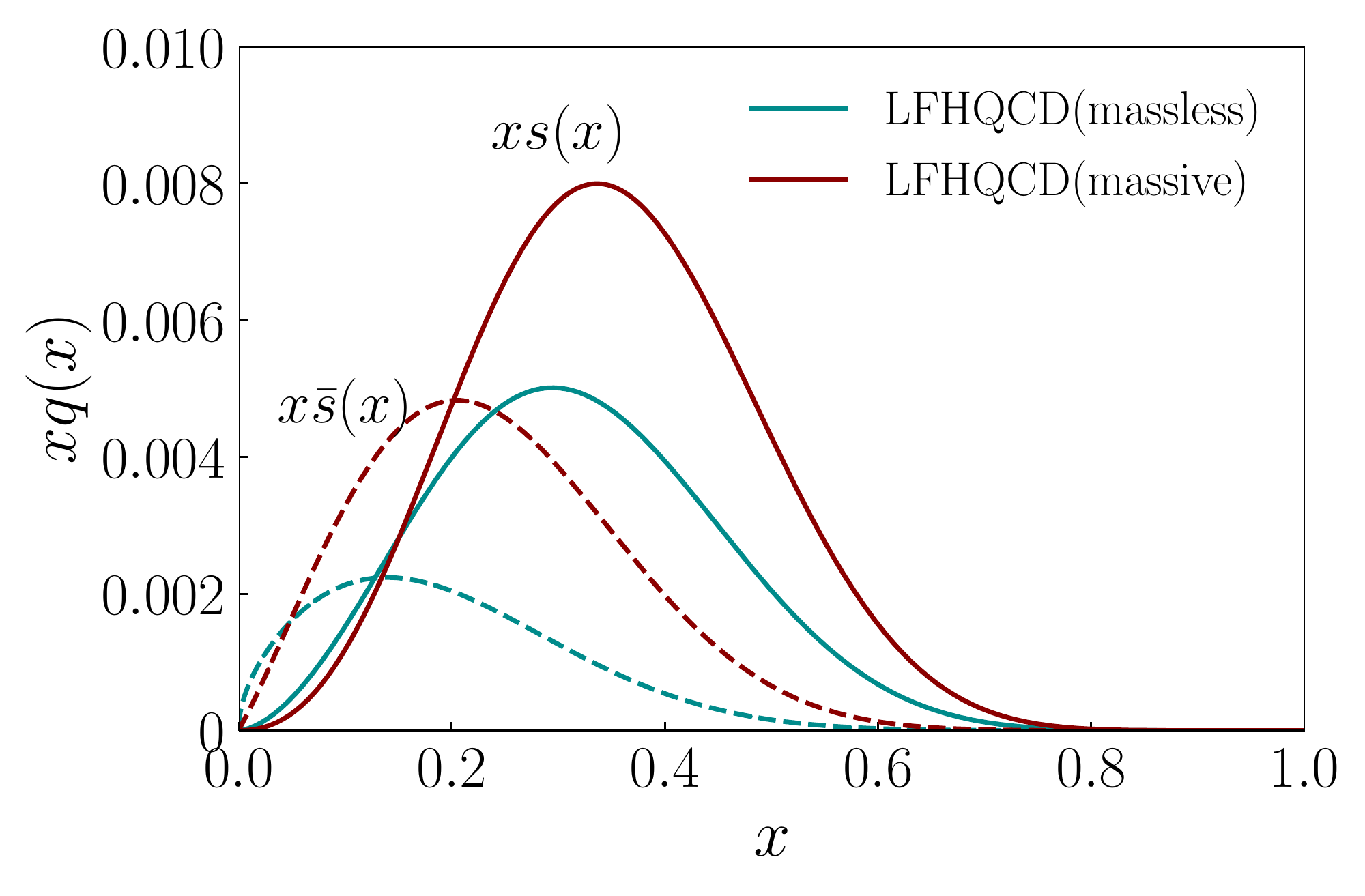}
\includegraphics[width=8.56cm]{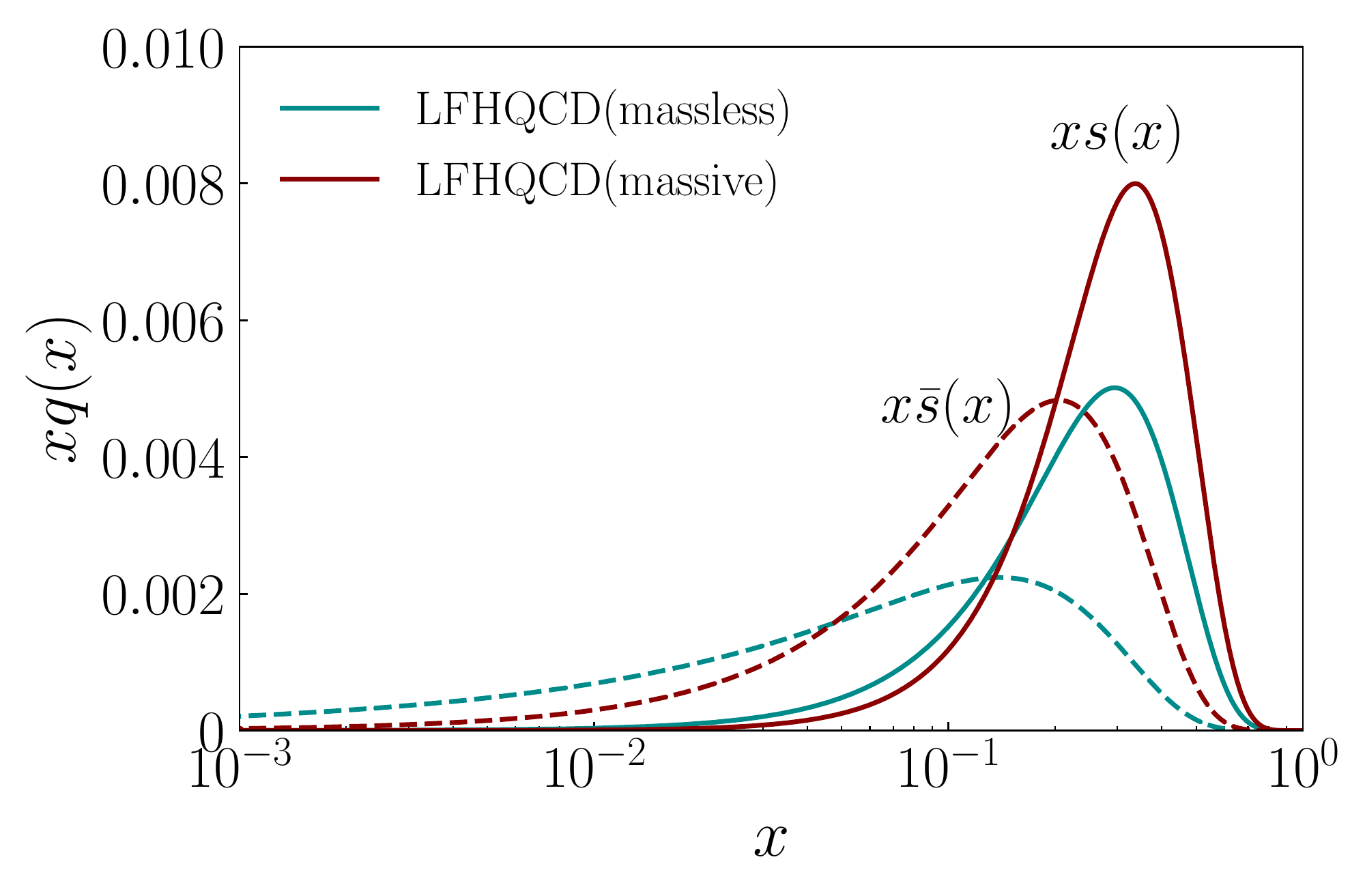}
\caption{The distributions $xs(x)$ (continuous curves) and $x\bar s(x)$ (dashed curves) correspond to the minimum intrinsic strange probability  $I_s = 0.2 \, N_s$ with $N_s = 0.047$, $\sqrt \lambda = 0.534\,\rm GeV$, and $M_\phi^2 = 1.96 \, \lambda$. The results with massless quarks are included for comparison.}
\label{sep}
\end{center}
\end{figure}

The solution which minimizes the  strange sea probability corresponds to $\alpha = N_s$ and  $I_s = (1-R)N_s$ with longitudinal quark distributions
\begin{eqnarray} \label{s}
s(x) &=&  N_s \, q_{\tau= 5}(x) + (I_s - N_s) \, q_{\tau= 6}(x) ,\\ \label{bs}
\bar s(x) &=&  I_s  \, q_{\tau =6}(x).
\end{eqnarray}
We show in Fig.~\ref{sep} the holographic results for the individual quark distributions $s(x)$ and $\bar s(x)$. The results correspond to the lower bound $I_s = 0.92 \%$. As we  discussed in Sec.~\ref{General}, the strange distribution $s(x)$ should have its support for larger values of the longitudinal momentum $x$, as compared with $\bar s(x)$, to lead to negative $s(x) - \bar s(x)$ asymmetry at small-$x$ and to a positive asymmetry at large-$x$. This important property is verified for the holographic quark distributions shown in Fig. \ref{sep}. One can observe in Fig.~\ref{sep} (left) that the high-twist suppression at large-$x$ from local counting rules is significant for the $s(x)$ leading-twist-5 distribution above $x \sim 0.7$ and for the $\bar s(x)$ twist-6 distribution above $x \sim 0.6$.

The positive form factor $F_1^s(Q^2)$ obtained from the lattice calculations~\cite{Sufian:2016pex, Sufian:2016vso}, shown in Fig.~\ref{f1sWF}, requires that the strange quarks are more concentrated at small transverse separation compared with the antistrange quarks (See Sec. \ref{General}). As  shown in Fig. \ref{rhoa} this is indeed the case for the  LFHQCD results computed from the coordinate space transverse distribution given by Eq.~\eqref{rhoaperp}.

\begin{figure}[htp] 
\begin{center} 
\includegraphics[width=0.49\textwidth]{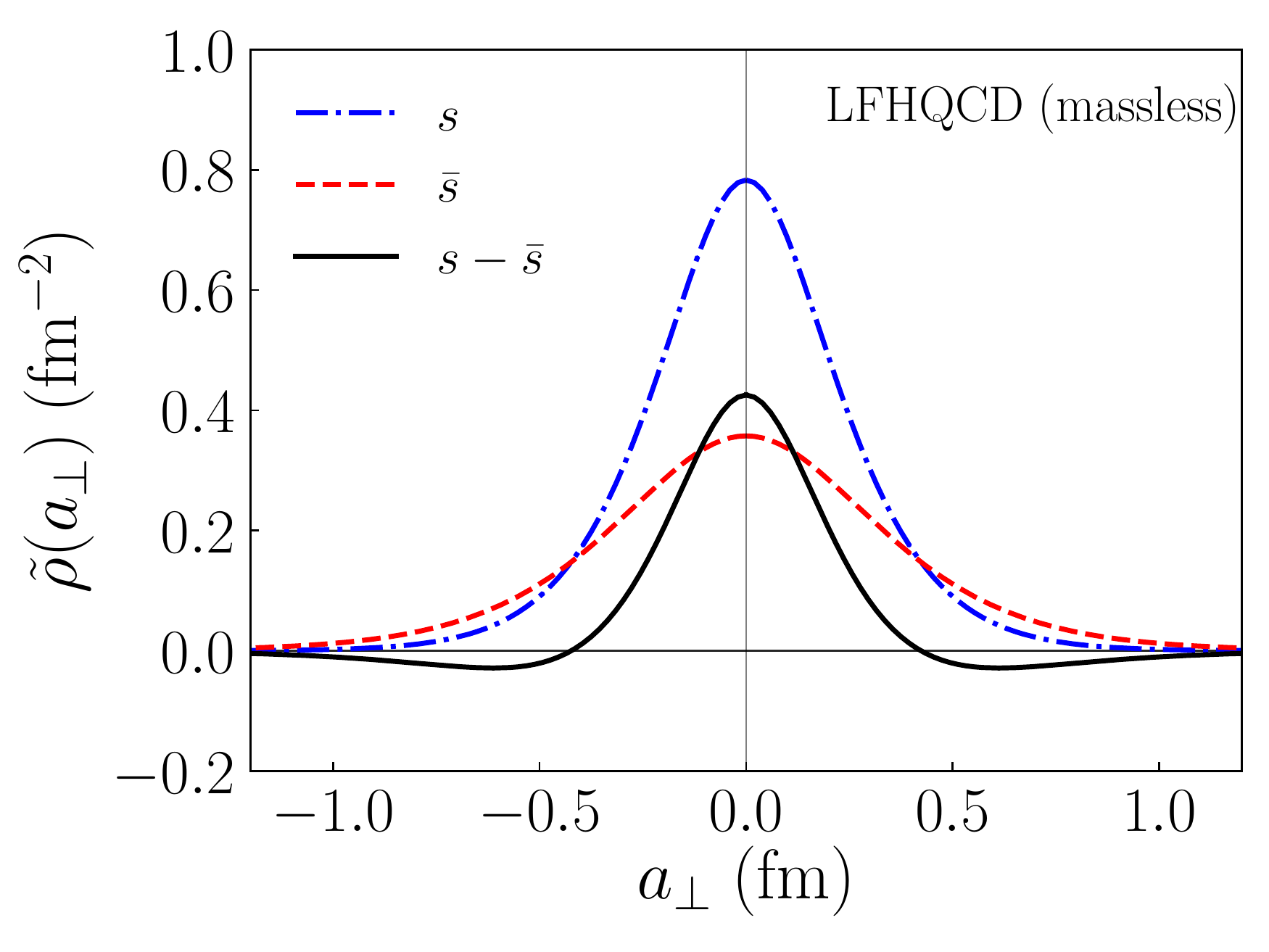}
\includegraphics[width=0.49\textwidth]{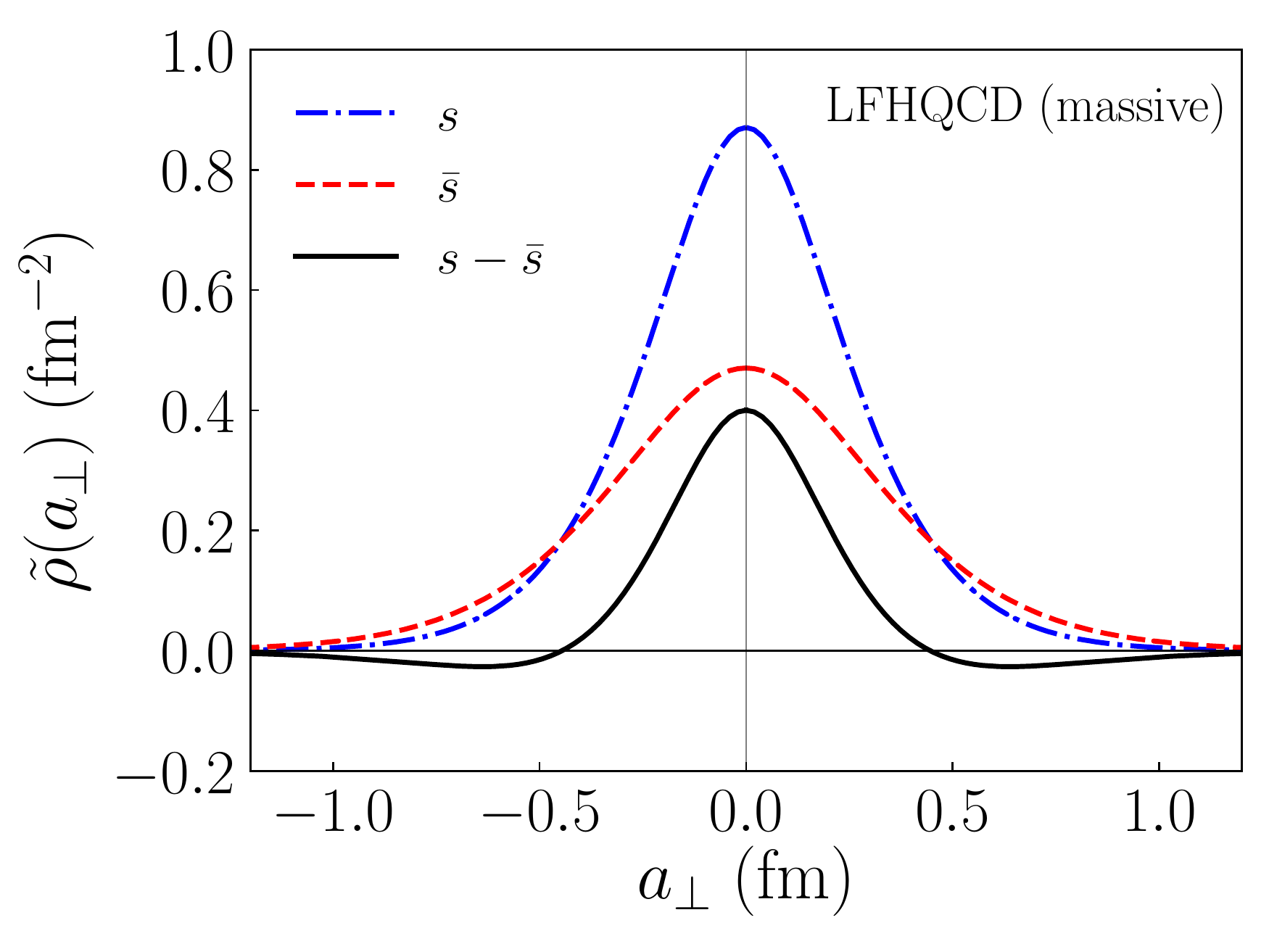}
\caption{Light-front holographic results for the asymmetric strange and antistrange quark distributions in transverse coordinate space corresponding to the minimum possible intrinsic strange probability.}
\label{rhoa}
\end{center}
\end{figure}


\section{\label{Concl} Discussions and conclusions}

In this article, we have demonstrated that a nonzero strangeness contribution to the spacelike electromagnetic form factor of the nucleon $F_1^s(Q^2) \ne 0$ implies a strange-antistrange asymmetry  in the nucleon's light-front wave function and thus in the nucleon PDF.

A lattice  QCD calculation predicts a positive strange quark form factor, which indicates that the strange quark distribution is more centralized in coordinate space than the antistrange quark distribution. Consequently, the strange quark distribution is more spread out in momentum space. The lattice result thus indicates a negative $s(x)-\bar{s}(x)$ longitudinal momentum distribution at small-$x$ and a positive distribution at large-$x$.

We have  shown how the baryon-meson fluctuation model leads to a nonzero  strange quark form factor of the  nucleon, and a $s(x)-\bar{s}(x)$ asymmetry. Imposing the lattice QCD data, we have analyzed the constraints on the model, leading to $1.06(51)\%$ intrinsic strange sea quark probability in the nucleon.

We have also discussed a new model for the intrinsic sea-quark distributions based on light-front holographic QCD. The strange quark form factor and the $s(x)-\bar{s}(x)$ asymmetry are determined in this framework up to a normalization factor, which can be constrained by the lattice prediction. Effects from the finite quark masses 
of the vector mesons which couples to the quark current in the nucleon have also been discussed. Remarkably, the holographic structure of form factors and PDFs allows the introduction of quark masses without modifying the hard scattering counting rules, the local counting rules,  or the $t$-dependence of GPDs. The small-$x$ behavior modified by quark masses is still governed by the Regge intercept. Since the strange quark mass is much greater than up and down quark masses, the strange quark distributions at small-$x$ in LFHQCD is less singular than up and down quark distributions.  By incorporating the positivity bound on quark distribution functions, we  have derived  a lower bound  for the intrinsic strange sea probability using the holographic approach.  The lower bound is 0.92\%, compatible with the value found in the fluctuation model; however, the intrinsic strangeness probability contributing to $s(x)+\bar{s}(x)$ can be significantly larger. We  have also evaluated the individual $s(x)$ and $\bar{s}(x)$ distributions and the coordinate-space transverse distributions for the strange and antistrange quarks in the  nucleon for the intrinsic strange quark probability determined by the lower bound. The result supports the qualitative analysis that the strange quark is more concentrated at small transverse separation than the antistrange quark.  This novel nonperturbative approach to sea quark distributions presented here, based on the light-front holographic framework, complements the physical picture inherent to the meson-baryon fluctuation model, and gives new insights into both the structure of the strange-antistrange asymmetry and the strange form factor of the nucleons.
This approach can also be extended to the study of intrinsic charm and bottom.

\begin{acknowledgments}

R.S.S. and T.L. thank Jian-Ping Chen and Wally Melnitchouk for useful discussions. This work is supported in part by the Department of Energy Contracts No. DE-AC05-06OR23177, No. DE-FG02-03ER41231, and No. DE-AC02-76SF00515. This work is also supported in part by National Natural Science
Foundation of China under Contracts No. 11775118 and No. 11475006.

\end{acknowledgments}


\appendix
\section{Lattice QCD determination of the strange quark form factor}
\la{LQCD}

The $s$ quark contribution to the nucleon's magnetic moment and charge radius has been calculated in Ref.~\cite{Sufian:2016pex} using the overlap fermion on the $(2+1)$ flavors RBC/UKQCD domain wall fermion (DWF) gauge configurations. Details of these ensembles are listed in Table~\ref{table1}. The authors used 24 valence quark masses in total for the 24I, 32I, 32ID, and 48I ensembles representing pion masses in the range $m_{\pi}\in$(135, 400) MeV to explore the quark-mass dependence of the strange quark form factors.

\begin{table}[htp]
\caption{\label{table1} The parameters for the DWF configurations: spatial/temporal size, lattice spacing \cite{Aoki:2010dy,Blum:2014tka}, the strange quark mass in the $\overline{\text{MS}}$ scheme at {2 GeV}, the pion mass corresponding to the degenerate light sea quark mass, and the numbers of configurations used in Ref.~\cite{Sufian:2016pex}.}
\begin{center}
\tabcolsep=0.03cm
\begin{tabular}{|c|c|c|c|c|c|}
\hline
~~~Ensemble~~~ & ~~~$L^3\times T$~~~  & ~~~$a$ (fm)~~~ & ~~~$m_s^{(s)}$ (MeV)~~~ &  ~~~{$m_{\pi}$} (MeV)~~~  & ~~~$N_\text{config}$~~~ \\
\hline
24I~\cite{Aoki:2010dy} & $24^3\times 64$& 0.1105(3) &120   &330  & 203    \\
\hline
32I~\cite{Aoki:2010dy} &$32^3\times 64$& 0.0828(3) & 110   &300 & 309 \\
\hline
32ID~\cite{Blum:2014tka} &$32^3\times 64$& 0.1431(7) & 89.4& 171 & 200\\
\hline
48I~\cite{Blum:2014tka} &$48^3\times 96$& 0.1141(2) & 94.9   &139 & 81 \\
\hline
\end{tabular}
\end{center}
\end{table}

One can perform the model-independent $z-$expansion fit to the form  factor $G(Q^2)$~\cite{Hill:2010yb,Epstein:2014zua}
\begin{align} \la{zexp}
G^{z-exp}(Q^2)=\sum^{k_{max}}_{k=0} a_k z^k, \qquad
z=\frac{\sqrt{t_{\text{cut}}+Q^2}-\sqrt{t_{\text{cut}}}}{\sqrt{t_{\text{cut}}+Q^2}+\sqrt{t_{\text{cut}}}},
\end{align} 
using the lattice data of strange Sachs electric and magnetic form factors $G^s_{E,M}(Q^2)$ to extrapolate the $s$-quark magnetic moment and charge radius as shown in~\cite{Sufian:2016pex}, and then use the fit parameters $a_k$ to interpolate $G^s_{E,M}$ values at various $Q^2$ for a given valence quark mass on the lattice. The available $Q^2$ on the 24I and 32I ensembles are $Q^2\in(0.22, 1.31)\,\text{GeV}^2$, on the 32ID ensemble are~$Q^2\in(0.07, 0.43)\,\text{GeV}^2$ and on the 48I ensemble are $Q^2\in(0.05, 0.31)\,\text{GeV}^2$. It is a common problem for lattice QCD calculations that the signal-to-noise ratio decreases as one reaches the physical pion mass. Lattice results of $G^s_{E,M}(Q^2)$ at the physical pion mass on the 48I ensemble~\cite{Blum:2014tka} is noisier compared to the $G^s_{E,M}(Q^2)$ obtained from the lattice ensembles with heavier pion masses. Although the largest available momentum transfer of the 24I and 32I ensembles is $Q^2\sim 1.3$ GeV$^2$, the largest momentum transfer available on the 48I ensemble is $Q^2~\sim 0.31$ GeV$^2$. We note that  the uncertainties in the extrapolation of the nucleon strange electromagnetic form factor become very large and the form factors are consistent with zero above $Q^2~\sim 0.7$ GeV$^2$ for the 48I ensemble and therefore the extrapolations of the 48I ensemble electromagnetic form factor data were constrained up to $Q^2=0.5$ GeV$^2$ in the global fit (a simultaneous fit in lattice spacing, volume and pion mass) in Ref.~\cite{Sufian:2016vso}. It is important to note that the lattice QCD estimate of $G^s_{E,M}(Q^2)$ in Ref.~\cite{Sufian:2016vso} is the most precise and accurate first-principles calculation of $s$-quark EMFFs to date. This is the only calculation at the physical pion mass where the quark mass dependence, as well as finite lattice spacing ($a$), volume corrections, and partial quenching effect (when the valence and sea quark masses are not the same in lattice QCD simulation) were considered.\\ 

After obtaining $Q^2$-dependence from the $z$-expansion fit to the lattice data, for a given $Q^2$ -value, we obtain 24 data points corresponding to different valence quark masses from 3 different lattice spacings and volumes and 4 sea quark masses including one at the physical point. We use the chiral extrapolation formula from Ref.~\cite{Hemmert:1999mr} and volume correction from Ref.~\cite{Tiburzi:2014yra}, yielding a global fit in different quark masses, lattice spacings, volumes of the strange quark Sachs electric form factor at a given $Q^2$. It is given by
\bea \la{gsefit}
G^s_E (m_\pi,&&\!\!\!\!\! m_K,m_{\pi,vs}, a, L) = A_0 + A_1 m_K^2 +A_2 m_\pi^2 \nn\\
&+&A_3m_{\pi,vs}^2 + A_4 a^2 + A_5\sqrt{L} e^{-m_\pi L},
\eea
where $m_{\pi}/m_K$ is the valence pion/kaon mass and $m_{\pi, vs}$ is
the partially quenched pion mass $m_{\pi, vs}^2 = 1/2(m_{\pi}^2 + m_{\pi, ss}^2)$ with $m_{\pi, ss}$ the pion mass corresponding to the sea quark mass. The $\chi^2/\text{d.o.f.}$ for different $Q^2$ global fits ranges between 0.7-1.13. For example, in the continuum limit, the global fit for $Q^2=0.25\,\text{GeV}^2$ provides the physical value of $G^s_E\vert_\text{phys}=0.0024(8)$, $A_1=0.58(30)$, $A_2=-0.29(15)$, $A_3=-0.003(9)$, $A_4=0.001(2)$, and $A_5=-0.001(3)$ with $\chi^2/\text{d.o.f.}=1.1$. One could also consider a $\log(m_K)$-term in the chiral extrapolation of $G^s_E$ as shown in~\cite{Hemmert:1999mr}, however our analysis shows that this term does not have any effect on the global fit for our lattice data. A similar vanishing difference has been observed if one considers $e^{-m_\pi L }$ instead of  a $\sqrt{L}e^{-m_\pi L }$ term in the volume correction, where $L$ is the finite box size of a lattice. For example, including the factor $\log(m_K)$ and $e^{-m_\pi L }$ instead of $\sqrt{L}e^{-m_\pi L }$, one obtains $G^s_E\vert_\text{phys}=0.0026$ in comparison with $G^s_E\vert_\text{phys}=0.0024$. We include these small effects in the systematics of the global fit results. We also assign a 20\% systematic uncertainty from the model-independent $z$-expansion interpolation coming from adding a higher order term $a_3$ while fitting the $G^s_E(Q^2)$ data. These uncertainties are added in quadrature to the systematics discussed in~\cite{Sufian:2016pex}. \\

Similarly, we calculate the strange Sachs magnetic form factor $G^s_M$ at a particular $Q^2$ using the global fit formula
\bea \la{gsmfit}
G^s_M (&&\!\!\!\!\! m_\pi,m_K,m_{\pi,vs}, a , L) =A_0 + A_1 m_\pi + A_2 m_K \nn \\
&+& A_3m_{\pi,vs}^2 + A_4 a^2 + A_5m_\pi(1-\frac{2}{m_\pi L}) e^{-m_\pi L },
\eea
where we have used a chiral extrapolation linear in $m_\pi$ and $m_{\text{loop}}=m_K$~\cite{Musolf:1996zv, Hemmert:1998pi, Hemmert:1999mr, Chen:2001yi}. For the volume correction we refer to Ref.~\cite{Beane:2004tw}. From the global fit formula~(\ref{gsmfit}), for example, in the continuum limit at $Q^2=0.25\,\text{GeV}^2$, we obtain $G^s_M\vert_\text{phys}=-0.018(4)$, $A_1=0.04(3)$, $A_2 = -0.18(12)$, $A_3=-1.27(84)$, $A_4=0.008(6)$, and $A_5=0.04(5)$ with $\chi^2/\text{d.o.f.}=1.13$. From the values of the parameters in the global fit formula~(\ref{gsmfit}), it is seen that the quark mass dependencies play an important role in calculating $G^s_M(Q^2)$ at the physical point. A 9\% systematic uncertainty from the model-independent $z-$expansion and an uncertainty from the empirical fit formula have been included as discussed in~\cite{Sufian:2016pex}. We obtain systematics from the global fit formula by replacing the volume correction by $e^{-m_\pi L }$ only and also by adding a $m_{\pi,vs}$ term in the fit and include the difference in the systematics of the global fit results.
 
More details about the lattice analysis can be found in Refs.~\cite{Sufian:2016pex,Sufian:2017osl}.

\section{\label{VMRT} The vector meson trajectories in LFHQCD}
The meson spectrum in LFHQCD is given by~\cite{Brodsky:2014yha, Brodsky:2016yod}
\bea \label{M2}
M^2 = 4 \lambda\left(n + \frac{L + J}{2}\right) + \Delta M^2[m_1, \, m_2],
\eea
where the squared mass shift $\Delta M^2[m_1, m_2]$ incorporates the effect from finite light quark masses. Following the procedure discussed in Refs.~\cite{Brodsky:2014yha, Brodsky:2016yod}, one can add a correction term of the invariant mass $\sum_i \frac{m_i^2}{x_i}$ to the LF kinetic energy in the LF Hamiltonian, and leave, as a first approximation, the  LF transverse potential unchanged.  The resulting LF eigenfunction is then modified by the factor $e^{-\frac{1}{2\lambda} \sum_i \frac{m_i^2}{x_i}}$ by performing a Lorentz frame-invariant substitution in the LFWF~\cite{Brodsky:2008pg}. This leads, for a hadron with two constituents of mass $m_1$ and $m_2$, to the   correction of the quadratic mass spectra by the term:
\bea \label{DM2m}
\Delta M^2[m_1,m_2] &=& \frac{1}{N} \int_0^1 dx \left(\frac{m_1^2}{x} +\frac{m_2^2}{1-x}\right) \,e^{-\frac{1}{\lambda} \left(\frac{m_1^2}{x} +\frac{m_2^2}{1-x}\right)},\nn \\
N_m&=&  \int_0^1 dx \, e^{-\frac{1}{\lambda} \left(\frac{m_1^2}{x} +\frac{m_2^2}{1-x}\right)},
\eea
where the $m_i$ are effective quark masses.

The longitudinal confinement dynamics in presence of quark masses has also been discussed in Refs.~\cite{Chabysheva:2012fe, Li:2017mlw}. In~\cite{Li:2017mlw} a specific longitudinal confinement potential is introduced by extending the transverse holographic potential while maintaining rotational invariance in the heavy quark limit. The approaches of Refs.~\cite{Brodsky:2014yha} and~\cite{Li:2017mlw}  lead to very similar results for the ground state distribution amplitudes.

For vector mesons in the lowest radial excitation one obtains from \eqref{M2}:
\begin{align} \label{vm-spectra}
M^2=4 \lambda (J-\half)+\Delta M^2[m_1,m_2]. 
\end{align}
from which one  deduces the Regge trajectory:
\begin{align} \label{traj}
\alpha(t) = \alpha' \, t +\alpha(0),  
\end{align}
with slope $\alpha'$ and intercept $\alpha(0)$ given by
\begin{align} \label{trajm}
\alpha'=\frac{1}{4 \lambda}, \quad \quad  \alpha(0)= \frac{1}{2} - \frac{1}{4 \lambda} \Delta M^2[m_1,m_2].
\end{align}

The QCD scale  $\sqrt{\lambda} = \kappa$ is determined from the spectra in all light hadronic channels and it is remarkably independent of the channel (mesonic and hadronic)~\cite{Brodsky:2016yod}. Its value  is $\sqrt{\lambda} = \kappa = 0.523$ GeV, with a standard deviation of $0.024$ GeV. Therefore,  for mesons consisting of light quarks the Regge slope is universal, $\alpha'= \frac{1}{4 \lambda}$. In contrast, the intercept $\alpha(0)$ depends on the effective quark masses, see \eqref{trajm}. Using the measured values of the pion and kaon masses one obtains from $M_\pi = \Delta M^2[m_q, m_{\bar q}]$ and $M_K = \Delta M^2[m_q, m_{\bar s}]$ the values $m_q=m_u = m_d= 46$ MeV and $m_s=357$ MeV for the effective quark masses of the light quarks~\cite{Brodsky:2014yha,Brodsky:2016yod}. With these values for the effective quark masses one obtains the intercept of the $\rho, \, \phi$ and $K^*$ trajectories  
\begin{eqnarray}
\alpha_\rho(0) &=& \frac{1}{2} - \frac{m_\pi^2}{4 \lambda} , \\ 
\alpha_\phi(0)  &=& \frac{1}{2} - \frac{\Delta M^2[m_s,m_{\bar s}]}{4 \lambda} , \\ 
\alpha_{K^*}(0) &=& \frac{1}{2} - \frac{m_K^2}{4 \lambda}.
\end{eqnarray}
Here it was taken into account that the   $\phi$ meson is nearly a pure $s\bar{s}$ state~\cite{Amsler:PDG}.

Using  the mass shift  Eq.~\eqref{DM2m}   we find $\Delta M^2[m_s, m_{\bar s}] / \lambda = 2.16 \pm 0.20$.
This value is  slightly larger  than the value  1.96  extracted from the combined spectral fit in Sec.~\ref{strangeFF}, but agrees with it even within the statistical errors.  As final values for the intercepts from LFHQCD we obtain the intercept values  $\alpha_\rho(0) =  0.482 \pm 0.002$, $\alpha_\phi(0) =  -0.04 \pm 0.05$ and $\alpha_{K^*}(0) = 0.275 \pm 0.020$, to be compared with the fitted values from the spectra  $\alpha_\phi(0) =  0.01$ and  $\alpha_{K^*}(0) = 0.273$.

 \end{document}